\documentclass[longauth]{aa}  
\pdfoutput=1
\usepackage{graphicx}
\usepackage{amsmath}
\usepackage{txfonts}
\usepackage{natbib}
\DeclareGraphicsExtensions{.pdf,.png,.jpg}
\bibpunct{(}{)}{;}{a}{}{,}

\begin{document} 

\title{Galactic Archaeology with asteroseismology and spectroscopy:\\ 
Red giants observed by CoRoT and APOGEE\thanks{The data described in Table B.1 are only available in electronic form at the CDS via anonymous ftp to cdsarc.u-strasbg.fr (130.79.128.5) or via {\tt http://cdsweb.u-strasbg.fr/cgi-bin/qcat?J/A+A/}.}}

\author{F. Anders\inst{1,2}, C. Chiappini\inst{1,2}, T. S. Rodrigues\inst{2,3,4}, 
A. Miglio\inst{5}, J. Montalb\'{a}n\inst{4}, B. Mosser\inst{6}, 
L. Girardi\inst{2,3}, M. Valentini\inst{1}, A. Noels\inst{7},
T. Morel\inst{7}, J. A. Johnson\inst{8}, M. Schultheis\inst{9}, 
F. Baudin\inst{10}, R. de Assis Peralta\inst{6}, S. Hekker\inst{11, 12}, 
N. Theme\ss l\inst{11, 12}, T. Kallinger\inst{13}, R. A. Garc\'{i}a\inst{14}, S. 
Mathur\inst{15},
A. Baglin\inst{6}, B. X. Santiago\inst{2,16}, M. Martig\inst{17}, I. 
Minchev\inst{1}, M. Steinmetz\inst{1}, L. N. da 
Costa\inst{2,18}, M. A. G. Maia\inst{2,18}, 
C. Allende Prieto\inst{19,20}, K. Cunha\inst{18}, T. C. Beers\inst{21}, C. Epstein\inst{8}, A. E. Garc\'{i}a P\'{e}rez\inst{19,20}, D. A. Garc\'{i}a-Hern\'{a}ndez\inst{19,20}, P. Harding\inst{22}, J. Holtzman\inst{23}, S. R. Majewski\inst{24}, Sz. M\'{e}sz\'{a}ros\inst{25, 26}, D. Nidever\inst{27}, K. Pan\inst{23,28}, M. Pinsonneault\inst{8}, R. P. Schiavon\inst{29}, D. P. Schneider\inst{30,31}, M. D. Shetrone\inst{32}, K. Stassun\inst{33}, O. Zamora\inst{19,20}, G. Zasowski\inst{34}
}

\authorrunning{Anders, Chiappini, Rodrigues et al.}    
\titlerunning{Galactic Archaeology with CoRoT and APOGEE}

\institute{Leibniz-Institut f\"ur Astrophysik Potsdam (AIP), An der 
Sternwarte 16, 14482 Potsdam, Germany\\
  \email{fanders@aip.de, cristina.chiappini@aip.de, 
thaise.rodrigues@oapd.inaf.it}
\and{Laborat\'orio Interinstitucional de e-Astronomia, - LIneA, 
Rua Gal. Jos\'e Cristino 77, Rio de Janeiro, RJ - 20921-400, Brazil}
\and{Osservatorio Astronomico di Padova -- INAF, Vicolo 
dell'Osservatorio 5, I-35122 Padova, Italy}
\and{Dipartimento di Fisica e Astronomia, Universit\`a di Padova, Vicolo dell'Osservatorio 2, I-35122 Padova, Italy}
\and{School of Physics and Astronomy, University of Birmingham, 
Edgbaston, Birmingham, B15 2TT, United Kingdom}
\and{LESIA, Observatoire de Paris, PSL Research University, CNRS, Universit\'{e} Pierre et Marie Curie, Universit\'{e} Denis Diderot, 92195 Meudon, France}
\and{Institut d'Astrophysique et de G\'{e}ophysique, All\'{e}e du 6 aout, 17 - 
Bat. B5c, B-4000 Li\`ege 1 (Sart-Tilman), Belgium}
\and{The Ohio State University, Department of Astronomy, 4055 McPherson 
Laboratory, 140 West 18th Ave.,
Columbus, OH 43210-1173, USA}
\and{Observatoire de la Cote d'Azur, Laboratoire Lagrange, CNRS UMR 
7923, B.P. 4229, 06304 Nice Cedex, France}
\and{Institut d'Astrophysique Spatiale, CNRS, Universit\'{e} Paris XI, 91405 
Orsay Cedex, France}
\and{Max-Planck-Institut f\"{u}r Sonnensystemforschung, Justus-von-Liebig-Weg 3, 
37077 G\"{o}ttingen, Germany}
\and{Stellar Astrophysics Centre, Department of Physics and Astronomy, Aarhus University, Ny Munkegade 120, 8000 Aarhus C, Denmark}
\and{Institut f\"{u}r Astronomie, Universit\"{a}t Wien, T\"{u}rkenschanzstr. 17, 
Wien, Austria}
\and{Laboratoire AIM, CEA/DRF – CNRS - Univ. Paris Diderot – IRFU/SAp, Centre de Saclay, 91191 Gif-sur-Yvette Cedex, France}
\and{Space Science Institute, 4750 Walnut Street Suite  205, Boulder CO 80301, 
USA}
\and{Instituto de F\'\i sica, Universidade Federal do Rio Grande do 
Sul, Caixa Postal 15051, Porto Alegre, RS - 91501-970, Brazil}
\and{Max-Planck-Institut f\"{u}r Astronomie, K\"{o}nigstuhl 17, D-69117 
Heidelberg, 
Germany}
\and{Observat\'orio Nacional, Rua Gal. Jos\'e Cristino 77, Rio de 
Janeiro, RJ - 20921-400, Brazil}
\and{Instituto de Astrofisica de Canarias,  C/ Vía Láctea, s/n, 38205, 
La Laguna, Tenerife, Spain}
\and{Departamento de Astrof\'{i}sica, Universidad de La Laguna (ULL), E-38206  La Laguna, Tenerife, Spain}
\and{Dept. of Physics and JINA-CEE: Joint Institute for Nuclear Astrophysics -- Center for the Evolution of the Elements, Univ. of Notre Dame, Notre Dame, IN 46530 USA} 
\and{Department of Astronomy, Case Western Reserve University, Cleveland, OH 44106, USA}
\and{New Mexico State University, Las Cruces, NM 88003, USA}
\and{Department of Astronomy, University of Virginia, PO Box 400325, Charlottesville VA 22904-4325, USA}
\and{ELTE Gothard Astrophysical Observatory, H-9704 Szombathely, Szent Imre herceg st. 112, Hungary}
\and{Department of Astronomy, Indiana University, Bloomington, IN 47405, USA}
\and{Dept. of Astronomy, University of Michigan, Ann Arbor, MI, 48104, USA}
\and{Apache Point Observatory, PO Box 59, Sunspot, NM 88349, USA}
\and{Astrophysics Research Institute, Liverpool John Moores University, IC2, Liverpool Science Park 146 Brownlow Hill Liverpool L3 5RF, UK}
\and{Department of Astronomy and Astrophysics, The Pennsylvania State University, University Park, PA 16802}
\and{Institute for Gravitation and the Cosmos, The Pennsylvania State University, University Park, PA 16802}
\and{Mcdonald Observatory, University of Texas at Austin, HC75 Box 1337-MCD, Fort Davis, TX 79734, USA}
\and{Vanderbilt University, Dept. of Physics \& Astronomy, VU Station B 1807, Nashville, TN 37235, USA}
\and{Johns Hopkins University, Dept. of Physics and Astronomy, 3701 San Martin Drive, Baltimore, MD 21210, USA}
}

\date{Received 17.08.2015; accepted 05.08.2016}

  \abstract{
With the advent of the space missions CoRoT and {\it Kepler}, it has recently become feasible to determine precise asteroseismic masses and relative ages for large samples of red giant stars. 

We present the CoRoGEE dataset -- obtained from CoRoT light curves for 606 red giants in two fields of the Galactic disc that have been co-observed for an ancillary project of the Apache Point Observatory Galactic Evolution Experiment (APOGEE).

We used the Bayesian parameter estimation code PARAM to calculate 
distances, extinctions, masses, and ages for these stars in a homogeneous analysis, resulting in relative statistical uncertainties of $\lesssim 2\%$ in distance, $\sim 4\%$ in radius, $\sim 9\%$ in mass and $\sim25\%$ in age. We also assessed systematic age uncertainties stemming from different input physics and mass loss.

We discuss the correlation between ages and chemical abundance patterns of field stars over a broad radial range of the Milky Way disc (5 kpc $<R_{\mathrm{Gal}}<$ 14 kpc), focussing on the [$\alpha$/Fe]-[Fe/H]-age plane in five radial bins of the Galactic disc.
We find an overall agreement with the expectations of pure chemical-evolution models computed before the present data were available, especially for the outer regions. 
However, our data also indicate that a significant fraction of stars now observed near and beyond the solar neighbourhood migrated from inner regions.

Mock CoRoGEE observations of a chemodynamical Milky Way disc model indicate that the number of high-metallicity stars in the outer disc is too high to be accounted for even by the strong radial mixing present in the model. The mock observations also show that the age distribution of the [$\alpha$/Fe]-enhanced sequence in the CoRoGEE inner-disc field is much broader than expected from a combination of radial mixing and observational errors. We suggest that a thick-disc/bulge component that formed stars for more than 3 Gyr may account for these discrepancies.

Our results are subject to future improvements due to a) the still low statistics, because our sample had to be sliced into bins of Galactocentric distances and ages, b) large uncertainties in proper motions (and therefore guiding radii), and c) corrections to the asteroseismic mass-scaling relation. The situation will improve not only upon the upcoming {\it Gaia} data releases, but also with the foreseen increase in the number of stars with both seismic and spectroscopic information. 
}


\maketitle

\section{Introduction}\label{intro}

To reconstruct the formation history of the Milky Way, 
one would ideally like to obtain precise and unbiased ages for thousands or 
millions of stars in all parts of our Galaxy. 
To date, this goal is still far beyond 
reach, at least until astrometric parallaxes from the {\it Gaia} 
satellite \citep{Perryman2001} and asteroseismic data from K2 \citep{Howell2014} 
and PLATO 2.0 \citep{Rauer2014} will become available.

A common work-around for this problem is to use relative ``chemical clocks'' 
provided by element abundance ratios \citep{Pagel1997, Matteucci2001}:
each star carries in its atmosphere the enrichment history 
of the gas from which it was formed, only minimally polluted by its own 
stellar evolution, and accessible through spectroscopy. By combining this wealth of 
information with kinematic properties of stellar populations 
in different Galactic environments, we can systematically unravel the 
importance of the various physical processes that led to the formation of the 
Milky Way as we see it today (``Galactic Archaeology''; \citealt{Freeman2002, Turon2008}). 

Still, age determinations provide crucial constraints on several astrophysical processes: For example, the ages of old halo stars can be 
used as a lower limit for the age of the Universe \citep{Hill2002}.
The Galactic age-metallicity relation (e.g., \citealt{Twarog1980, Edvardsson1993, Ng1998}), the star-formation history \citep{Gilmore1999} or the 
evolution of abundance gradients (e.g., \citealt{Carraro1998, 
Chen2003}) are essential tools for understanding the evolution of our Milky Way. 

During the past ten years, ever more sophisticated chemodynamical models of Milky-Way-mass galaxies have been developed in a cosmological context (e.g., \citealt{Abadi2003, Stinson2010, Guedes2011, Brook2012a, Scannapieco2015, Roca-Fabrega2016}). However, detailed models that match many of the Milky Way's chemo-dynamical correlations \citep{Minchev2013, Minchev2014} are still rare (see discussions in \citealt{Scannapieco2012} and \citealt{Minchev2013}).
These can be compared to observations, but it is often difficult to find observables that are powerful enough to discard certain scenarios of the Galaxy's evolution. With the availability of age estimates for large stellar samples -- even if they are only valid in a relative sense -- this situation changes drastically.

It is therefore important to revisit the full age--chemistry--kinematics space with samples that cover larger portions of the Galactic disc. In this high-dimensional space, we can then look for robust statistical relations that realistic models have to fulfil. With the joint venture of asteroseismology and spectroscopic surveys, we are now in a position to constrain key parameters of stellar and Galactic evolution.


Unlike stellar radii and masses, the ages of stars cannot be directly 
measured, only inferred through modelling.
Among the various available stellar age indicators (e.g., Li abundance, U/Th 
ratio, stellar activity, rotation, X-ray luminosity, and position in the 
Hertzsprung-Russell diagram), 
one of the most promising methods that can deliver reliable age 
estimates for a wide range of ages is the comparison of measured 
atmospheric and asteroseismic parameters of evolved stars with models of 
stellar evolution \citep[e.g.,][]{Miglio2012}.

It is well-known (e.g., \citealt{Ulrich1986, 
Christensen-Dalsgaard1988}) that detailed asteroseismic 
analyses involving individual oscillation frequencies may deliver 
precise age determinations.
Depending on the spectral type of the star, a number of seismic 
characteristics can be used to investigate the stellar interior and infer an 
age estimate.
However, this so-called ``boutique'' or ``\`a la carte modelling'' 
\citep{Soderblom2013, Lebreton2014} requires 
extremely accurate measurements of several pulsation modes. To date, this is only possible for the Sun (e.g., \citealt{Gough2001}) and a relatively small number of bright dwarf stars observed by CoRoT and {\it Kepler} (e.g., \citealt{Metcalfe2010, 
Batalha2011, Mathur2012a, SilvaAguirre2013, Chaplin2013, Lebreton2014, Metcalfe2014}).

For large samples of red giant stars (first-ascent red giants as well as red-clump stars), statistical studies follow a different approach called ``ensemble asteroseismology'' \citep[e.g.,][]{Chaplin2011}. This method typically focusses on two main seismic characteristics of the frequency spectrum of solar-like oscillating giants: the large frequency separation $\Delta \nu$, related to the stellar mean density \citep{Tassoul1980, Ulrich1986, Christensen-Dalsgaard1993}, and the frequency of maximum oscillation power $\nu_{\mathrm{max}}$, related to the acoustic cut-off frequency 
\citep{Brown1991, Kjeldsen1995, Belkacem2011}. The mass and radius of a star have been shown to scale with these quantities via:
\begin{equation}
\begin{aligned}
 \dfrac{M}{M_{\odot}} &\simeq \Big(\frac{\nu_{\mathrm{max}}}{\nu_{\mathrm{max, 
\odot}}}\Big)^3 \Big(\frac{\Delta\nu}{\Delta\nu_{\odot}}\Big)^{-4} 
\Big(\frac{T_{\mathrm{eff}}}{T_{\mathrm{eff,\odot}}}\Big)^{3/2}, \\ 
 \dfrac{R}{R_{\odot}} &\simeq \Big(\frac{\nu_{\mathrm{max}}}{\nu_{\mathrm{max, 
\odot}}}\Big) \Big(\frac{\Delta\nu}{\Delta\nu_{\odot}}\Big)^{-2} 
\Big(\frac{T_{\mathrm{eff}}}{T_{\mathrm{eff,\odot}}}\Big)^{1/2},
\end{aligned}
\label{scale}
\end{equation}
where $T_{\mathrm{eff}}$ is the star's effective temperature, and the 
solar values $\Delta\nu_{\odot}=135.03\,\mu$Hz, $\nu_{\mathrm{max,\odot}}=3140.0 \,\mu$Hz, and $T_{\mathrm{eff, \odot}} = 5780$ K \citep{Pinsonneault2014} are used in the following.

The scaling relations (\ref{scale}) have been tested with independent methods
in the past years (eclipsing binaries, open clusters, interferometry, {\it Hipparcos} parallaxes), and shown to be valid for a broad parameter regime 
(see \citealt{Chaplin2013} for a review). Possible systematic biases concerning the mass determination are introduced by departures from a simple scaling of $\Delta \nu$ with the square root of the stellar mean density (see e.g., \citealt{White2011a, Miglio2012, Miglio2013a, Belkacem2013a}). Suggested corrections to the $\Delta \nu$ scaling probably depend (to a level of a few percent) on the stellar structure itself. Moreover, the average $\Delta \nu$ is known to be affected (to a level of around 1\% in the Sun) by inaccurate modelling of near-surface layers.

The seismic mass of a red giant provides a 
powerful constraint on its age, because its red-giant branch (RGB) lifetime is relatively short compared to its main-sequence lifetime. 
Combined with independent measurements of metallicity and 
effective temperature, the main seismic characteristics provide good statistical measures for the primary derived parameters of a star, 
such as mass, radius, distance, and age (e.g., \citealt{Miglio2012, Rodrigues2014, Casagrande2016}).

Unfortunately, the overall quality (in terms of precision as well as accuracy) of age determinations for giant stars is still fairly limited (e.g., \citealt{Jorgensen2005, Soderblom2010, Casagrande2016}). Systematic age uncertainties depend on the quality of the observables along with their uncertainties, as well as theoretical uncertainties of stellar models (e.g., \citealt{Noels2015}; see also Sect. \ref{sys}).

With the recently established synergy of asteroseismology and high-resolution 
spectroscopy surveys, it has become possible to determine more 
precise ages for red giants.


\begin{figure*}
\centering
  \includegraphics[width=0.9\textwidth]{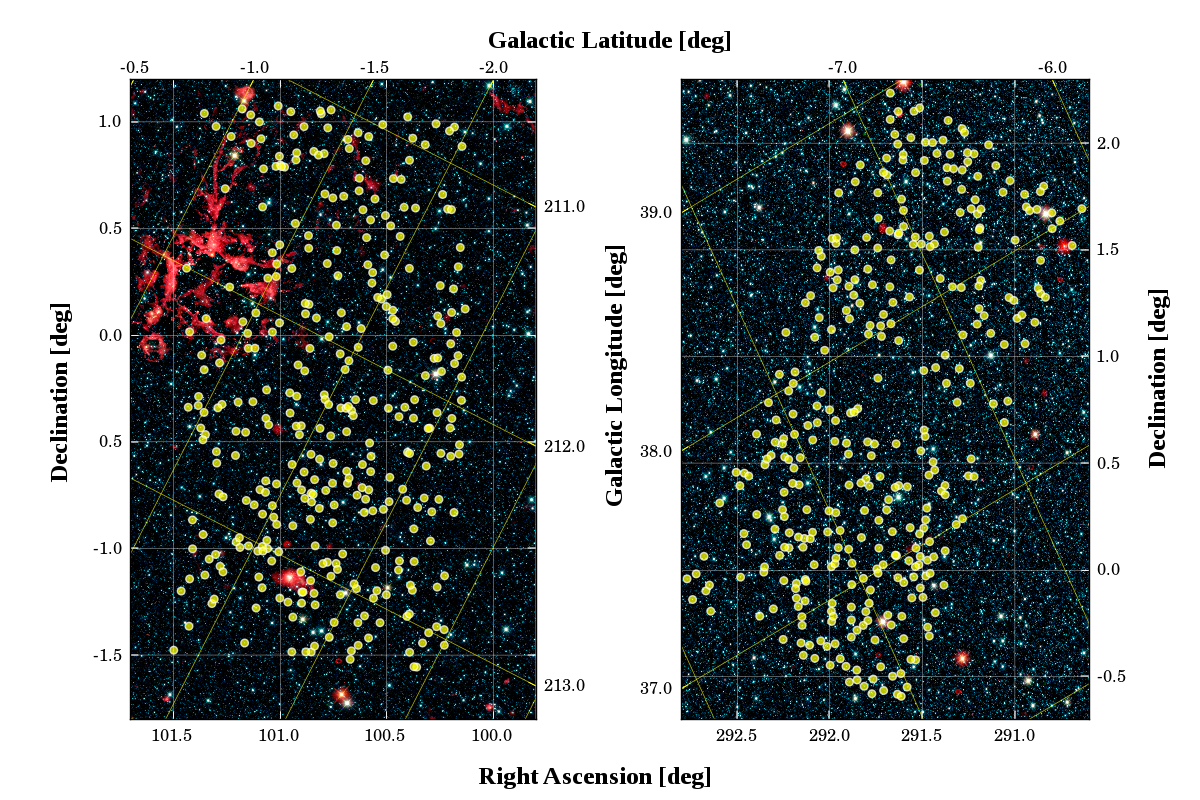}
    \caption{Location of the stars observed with APOGEE in the two CoRoT 
exoplanet fields LRa01 ({\it left}) and LRc01 ({\it right}). Indicated 
in yellow are the stars for which asteroseismic parameters were available. The 
background colour image is composed of near-infrared WISE {\it W1, W2} and {\it W3} images from the AllWISE data release \citep{Cutri2013}. Bottom and outside vertical labels display equatorial coordinates, while the top and central vertical labels refer to Galactic coordinates. Corresponding coordinate grids are also shown.}
   \label{fieldmaps}
\end{figure*}

The detection of solar-like oscillations in thousands of field stars by CoRoT and {\it Kepler} has opened the door to detailed studies of the Milky Way's stellar populations. Data from the first CoRoT observing run revealed solar-like oscillations in thousands of red giants \citep{Hekker2009}. \citet{Miglio2009} presented a first comparison between observed and predicted seismic properties of giants in the first CoRoT field, which highlighted the expected signatures of red-clump stars in the $\Delta \nu$ and $\nu_{\mathrm{max}}$ distributions. 
\citet{Miglio2013a} presented a first comparison between populations of red giants observed by CoRoT in two different parts of the Milky Way (the CoRoT fields LRa01 and LRc01 also investigated here; see Fig. \ref{fieldmaps}), which showed significant differences in the mass distributions of these two samples, and were interpreted as mainly due to the vertical gradient in the distribution of stellar masses (hence ages) in the disc (see also \citealt{Casagrande2016} for a first measurement of the vertical disc age gradient).
However, the precision of the age determinations used in this pilot study was still limited to $30-40\%$, due to the absence of constraints on 
photospheric chemical composition \citep{Miglio2013}. 

Recently, large-scale follow-up observations of seismic targets have begun. The SAGA project \citep{Casagrande2014b, Casagrande2016} is covering the {\it Kepler} field with Str\"{o}mgren photometry, thereby obtaining more precise stellar parameters. Similarly, spectroscopic stellar surveys such as RAVE \citep{Steinmetz2006}, APOGEE \citep{Majewski2015}, the Gaia-ESO survey \citep{Gilmore2012}, LAMOST \citep{Zhao2012}, and GALAH \citep{Zucker2012} are observing CoRoT and {\it Kepler} targets to anchor their spectroscopic surface gravity and distance measurements \citep[e.g., ][]{Bovy2014b, Holtzman2015} -- and to ultimately use the combined datasets to constrain the chemodynamical evolution of the Milky Way. The CoRoT-APOGEE (CoRoGEE) dataset paves the way for future advances in this direction.

Our paper is structured as follows: the CoRoGEE sample and the provenance of the different data (asteroseismology, spectroscopy, photometry and astrometry) are presented in Sect. \ref{obs}. Section \ref{ana} summarises our analysis and leads to our estimates of the main stellar \frqq desirables\flqq, such as mass, radius, age, distance, extinction, and kinematical parameters. 
We emphasise that our age estimates should be considered relative age indicators that are to be used in a statistical sense only.

\begin{figure}\centering
 \includegraphics[trim=0 200 0 0, clip, width=0.49\textwidth]{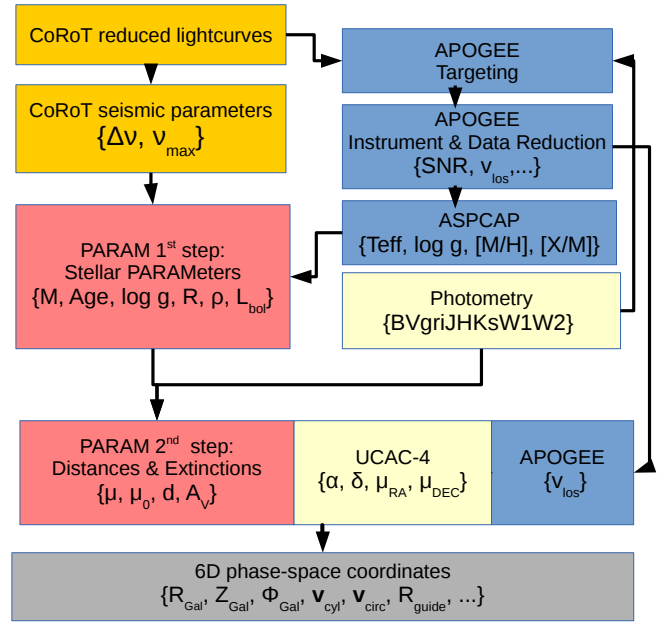}
 \caption{Overview of the data provenance and analysis steps performed for the 
CoRoT-APOGEE (CoRoGEE) data. Blue boxes correspond to APOGEE data products, 
orange boxes to CoRoT data, and light yellow boxes to existing catalogue data. 
Red boxes summarise the two parts of the PARAM pipeline, while the grey box 
summarises the kinematical data used for this work.}
 \label{dataflow}
\end{figure}

In Sect. \ref{res}, we use our sample to study for the first time the variation of the [$\alpha$/Fe]-vs.-[Fe/H]\footnote{The abundance ratio of two chemical elements X and Y is defined as $ \mathrm{[X/Y]} = \lg \frac{n_X}{n_Y} - \lg(\frac{n_X}{n_Y})_{\odot} $, 
where $n_X$ and $n_Y$ are respectively the numbers of nuclei of elements X and Y, per unit volume in the stellar photosphere.} relation with Galactocentric distance in three broad age bins, and compare our data with predictions from a chemodynamical Galaxy model. We conclude and discuss future paths to improve our analysis in Sect. \ref{conclu}.

The CoRoGEE dataset covers a wide radial range of the Galactic 
disc and provides precise stellar parameters, distances, and chemical abundances. Therefore, the presented data provide material for a number of subsequent analyses. In two companion  papers, we focus on specific results: 1. the discovery of an apparently young stellar population with enhanced [$\alpha$/Fe] ratio \citep{Chiappini2015a}, and 2. the variation of the disc's radial metallicity profile with stellar age (Anders et al., subm. to A\&A). The data are publicly available at the CDS (see online Appendix B).

\section{Observations}\label{obs}

\begin{figure*}\centering
\includegraphics[width=.9\textwidth]
{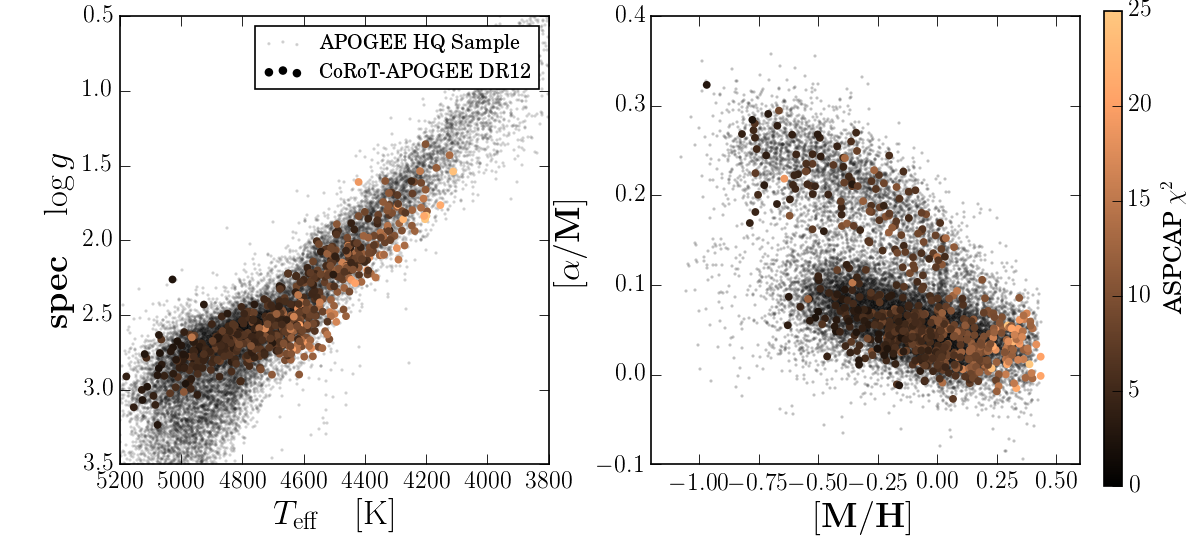} 
 \caption{Location of the CoRoT-APOGEE stars in the ASPCAP $\log 
g$-vs.-$T_{\mathrm{eff}}$ {\it Kiel} diagram (left) and the 
[$\alpha$/Fe]-vs.-[Fe/H] chemical abundance plane. The colour encodes 
the reduced $\chi^2$ of the ASPCAP fit. In the background, we plot the 
APOGEE DR10 high-quality giant sample \citep{Anders2014}, updated to DR12 atmospheric parameters, for comparison.}
 \label{aspcap}
\end{figure*}
 
Our observations combine the global asteroseismic parameters derived from 
precision light curves obtained by the CoRoT satellite \citep{Baglin2006, 
Michel2008} with stellar parameters and chemical abundances inferred from 
near-infrared (NIR) high-resolution spectra taken by the Apache Point 
Observatory Galactic Evolution Experiment (APOGEE). The field maps of the two CoRoT fields observed with APOGEE are shown in Fig. \ref{fieldmaps}. An overview on the data assembly and analysis is given in Fig. \ref{dataflow}.

\subsection{Adopted seismic parameters}\label{seismo}

The CoRoT data used in this work are a subset of the data analysed by 
\citet{Mosser2010} and \citet{Miglio2013a}: The CoRoT long runs in the LRa01 
and LRc01 exoplanet fields comprise photometric time series for several 
thousand stars of about 140 days, resulting in a frequency resolution of 
$\sim 0.08~\mu$Hz. For stars with detectable solar-like 
oscillations, \citet{Mosser2010} determined the large 
frequency separation, $\Delta \nu$, and the frequency of maximum oscillation 
power, $\nu_{\mathrm{max}}$, from the frequency spectra with the envelope autocorrelation-function method \citep{Mosser2009}, but without reporting individual uncertainties for these quantities.

In the following, we use the seismic parameters obtained from CoRoT N2 light curves\footnote{\url{http://idoc-corot.ias.u-psud.fr/jsp/doc/DescriptionN2v1.3.pdf}} in the same way as in \citet{Mosser2010}, updated to deliver individual uncertainties on $\Delta \nu$ and $\nu_{\mathrm{max}}$. When the envelope autocorrelation signal is high enough, a more precise estimate of the large separation is provided by the use of the so-called universal pattern method \citep{Mosser2011}. A comprehensive data release of newly reduced CoRoT light curves and higher-level science products, using analyses of several different seismic pipelines, will be presented in a separate paper.

As shown in \citet{Mosser2010} and \citet{Miglio2013a}, the target selection 
for the CoRoT asteroseismology program is homogeneous in both fields: 
solar-like oscillations were searched for in giant stars obeying the following 
cuts in the colour-magnitude diagram: $K_s<12, 0.6<J-K_s<1.0$. 
\citet{Mosser2010} also demonstrated that, for a wide parameter range, the selection 
bias introduced by the additional requirement of detected oscillations does 
not measurably affect the $\Delta \nu$ or $\nu_{\mathrm{max}}$ distributions in the 
two fields. 

\subsection{Spectroscopic data}\label{spec}

\begin{figure*}\centering
\includegraphics[width=.9\textwidth]
{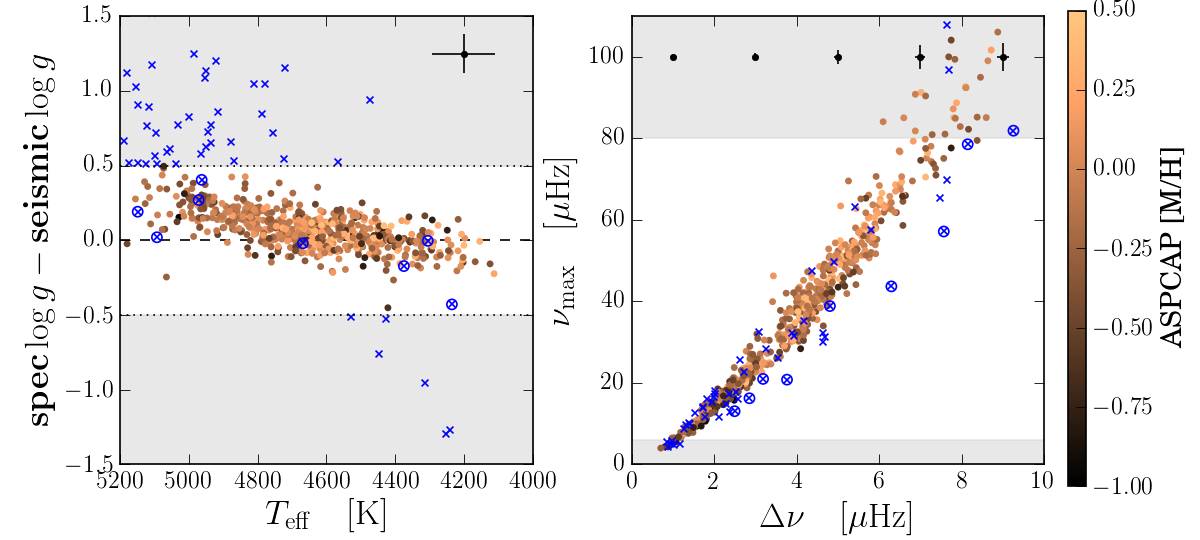}
 \caption{Left: Difference between ASPCAP (calibrated) $\log g$ and gravity 
determined from seismic scaling relations, as a function of effective 
temperature. Stars located in the grey-shaded area ($|\log 
g_{\mathrm{ASPCAP}} - \log g_{\mathrm{seismo}}|>0.5$ dex; {\it blue crosses}) were excluded from the analysis. 
Right: $\nu_{\mathrm{max}}-\Delta\nu$ diagram for our program stars. In 
addition to the $\log g$ consistency requirement, nine stars (mostly located far from the $\nu_{\mathrm{max}}-\Delta\nu$ sequence; {\it blue crossed circles}) were also rejected by the PARAM pipeline. 
Error bars in the upper part of the diagrams represent average uncertainties.}
 \label{aspcapseismo}
\end{figure*}

APOGEE \citep{Majewski2015} is a Galactic Archaeology experiment operating during the third and fourth epochs of the Sloan Digital Sky Survey (SDSS-III, 
\citealt{Eisenstein2011}; SDSS-IV). It uses the 
2.5 m telescope at APO \citep{Gunn2006} to feed a multi-object NIR fiber 
spectrograph \citep{Wilson2010, Wilson2012} that delivers high-resolution 
($R\sim 22,500$) 
H-band spectra ($\lambda=1.51-1.69~\mu$m) of 
mostly red giants. Dedicated processing and analysis pipelines 
(\citealt{Nidever2015, Holtzman2015}) allow for 
the determination of precise ($\sim100$ 
m/s) and accurate ($\sim 350$ m/s) radial velocities. In addition, the APOGEE Stellar Parameter and Chemical Abundances Pipeline  (ASPCAP; Garc\'ia P\'erez et al. 2015) provides stellar parameters and elemental abundances of 15 chemical elements from the best fit over extensive grids of pre-calculated synthetic stellar spectra \citep{Zamora2015} to the observed spectra. 

As an SDSS-III/APOGEE ancillary program, 690 
stars with detected seismic oscillations in the two CoRoT exoplanet fields 
LRa01 (APOGEE fields COROTA and COROTA3; $(l,b)_{\mathrm{cen}}=(212,-2)$) and 
LRc01 (COROTC; $(l,b)_{\mathrm{cen}}=(37,-7)$) were observed with the APOGEE 
instrument, at high signal-to-noise ratios 
(median $S/N$ of 230 per resolution element). The field maps of the observed targets are shown in Fig. \ref{fieldmaps}. The APOGEE targeting scheme allows 
for the combination of spectra taken at different times, so-called visits. 
Most of the stars ($\sim80\%$) have been observed at least three times
to reach the signal-to-noise ratio goal of 100, which is necessary to infer precise chemical abundance information \citep{Zasowski2013}. 

Unfortunately, the actual target selection for APOGEE observations of CoRoT solar-like oscillating red giants has not been carried out on the basis of a simple selection function.
The targets on the plates observed by APOGEE are a mixture of: 
\begin{enumerate}
 \item solar-like oscillating stars identified by \citet{Mosser2011} -- preferentially selected to be RGB stars,
 \item CoRoT stars observed by the Gaia-ESO survey \citep{Gilmore2012} for the purpose of cross-calibration, and
 \item APOGEE main-survey targets that were found to show solar-like oscillations in CoRoT, but were not selected on that basis.
\end{enumerate}
Therefore, the best way to correct for the CoRoGEE selection function is to compare what was observed with what could have been observed (i.e., compare the resulting spectro-seismic sample with the underlying photometric sample). 
In addition, it is necessary to assess whether the photometric parent sample (red giants in the fields LRa01 and LRc01) is representative of the overall stellar content in these fields (as done in \citealt{Miglio2013a, Miglio2013}). 
Both steps can be accomplished with stellar population synthesis modelling (see \citealt{Anders2016a}). One intermediate selection effect that we cannot address with the current CoRoGEE sample is whether the red giants with detected solar-like oscillations are fully representative of the underlying population. For the {\it Kepler} field, \citet{Casagrande2016} found that this is only true for a narrower region in the colour-magnitude diagram than we are considering here; our giant sample may therefore be slightly biased against redder colours (more evolved stars).

For this work, we make use of the ASPCAP-derived stellar parameters 
effective temperature, $T_{\mathrm{eff}}$, scaled-solar metallicity, [M/H], and 
relative $\alpha$-element abundance, [$\alpha$/M], from the SDSS data release 12 \citep[DR12][]{Alam2015, Holtzman2015}\footnote{We estimate the 
uncertainties in these abundances as $\sigma\mathrm{[M/H]} = 
\sigma\mathrm{[Fe/H]}$ and $\sigma\mathrm{[\alpha/M]} = 
\sqrt{\sigma\mathrm{[Mg/H]}^2 + \sigma\mathrm{[Fe/H]}^2}$.}. For 
the comparison to stellar isochrones, we approximated the overall metal 
abundance by the sum [Z/H] $\simeq$ [M/H]$_{\mathrm{uncalib}}$ 
$+[\alpha$/M]$_{\mathrm{uncalib}}$ (e.g., \citealt{Salaris1993, Anders2014}).
Fig. \ref{aspcap} summarises the distribution of the CoRoGEE stars in ASPCAP parameter 
space. We used calibrated values for the ASPCAP $T_{\mathrm{eff}}$ and surface gravity $\log g$.

To ensure that the ASPCAP stellar parameters and chemical abundances 
do not suffer from unknown problems, we discarded 12 stars that did not
satisfy the high-quality criteria laid out in \citet{Anders2014}. We also flagged and removed 14 stars for which a visual inspection of the CoRoT light curves revealed spurious detection of solar-like oscillations.
In addition, we required that the difference between the spectroscopically derived surface gravity be not too far from the value predicted by the seismic scaling relations: $|\log g_{\mathrm{ASPCAP}}^{\mathrm{calib}} - \log 
g_{\mathrm{seismo}}|<0.5$ dex. This criterion removed 47 stars for which the ASPCAP solution is incompatible with the seismic measurements (crosses in Fig. \ref{aspcapseismo}, left panel). In 
addition, 11 stars were rejected by our stellar parameter pipeline because 
their measured input values $\{\Delta\nu$, $\nu_{\mathrm{max}}, T_{\mathrm{eff}},$ [M/H]\} were incompatible with any stellar model within their uncertainties (crossed circles in Fig. \ref{aspcapseismo}, right panel).

\begin{figure*}\centering
 \includegraphics[width=.99\textwidth]
 {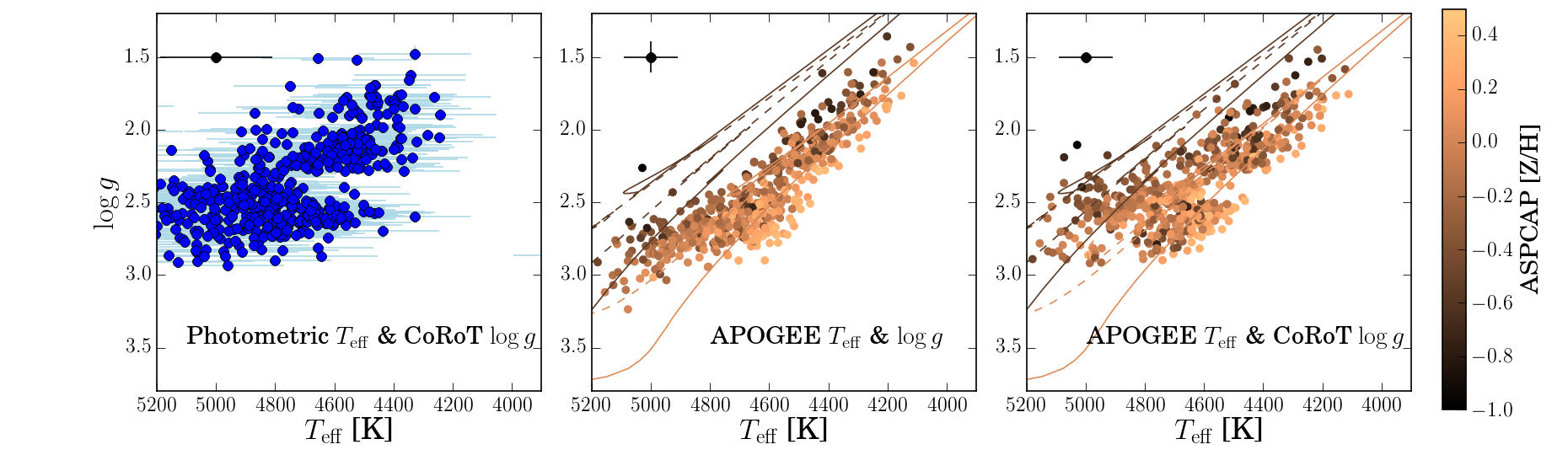}
 \caption{$\log g - T_{\mathrm{eff}}$ {\it Kiel} diagrams. Left: 
Photometric temperatures and $\log g$ from CoRoT seismic parameters + scaling 
relations. Middle: Purely spectroscopic diagram using APOGEE stellar 
parameters, colour-coded by metallicity. Right: Joint CoRoT-APOGEE {\it Kiel} 
diagram. Overplotted are PARSEC isochrones \citep{Bressan2012} for 
metallicities $-0.6$ and $0.0$ at ages 1.4 (dashed lines) and 4.5 Gyr (solid 
lines) for comparison. 
As noted by \citet{Martig2015}, there is 
a slight temperature 
discrepancy between models and data for 
sub-solar metallicities. Error bars in the upper left of each panel indicate 
median uncertainties.}  
 \label{kieldiagrams}
\end{figure*}

\subsection{Photometry and astrometry}\label{phot}
To determine distances to the stars in our sample with the best 
possible precision, the spectroscopic and asteroseismic information was 
complemented by photometric data obtained over a wide wavelength range. 

Standard Harris $B$ and $V$ as well as Sloan-Gunn $r'$ and $i'$ magnitudes are 
available for our CoRoT targets from the OBSCAT catalogue which was released as 
a supplement to the EXODAT archive (\citealt{Meunier2007}, Damiani et al., in prep.). The observations were performed with the Wide Field Camera (WFC) at the 2.5 m Isaac Newton Telescope (INT) at Roque de los Muchachos Observatory (La Palma) in 2002\footnote{\url{http://cesam.oamp.fr/exodat/index/exodat-documentation\#Photometryavailableforsubsamples}}.
 
Because the photometry of the USNO-B catalogue (which is also provided by 
EXODAT) is based on digitised photographic Schmidt plates and its calibration 
suffers from inaccuracies and inhomogeneities of about 0.2 
mag \citep{Monet2003}, we refrained from using this database.

We also added Johnson $BV$ and Sloan $g'r'i'$ photometry from the 
APASS survey's 6$^{\mathrm{th}}$ data release \citep{Henden2014}, with 
photometric accuracies of about 0.02 mag. 

In the infrared, accurate $JHK_s$ photometry is available from the
2MASS Point Source Catalog \citep{Cutri2003}, which served as the major input 
catalogue for APOGEE. We also added WISE $W1W2$ filters from the 
AllWISE Catalog \citep{Cutri2013} for which the photometric precision is sufficient to constrain the mid-infrared region of the stellar spectral energy 
distribution\footnote{As in \citet{Rodrigues2014}, we discard the filters $W3$ 
and $W4$ because of possible contamination by warm interstellar dust 
(e.g., \citealt{Davenport2014}) and larger measurement uncertainties.}.

For kinematical studies, proper motions were compiled from the recent UCAC-4 
catalogue \citep{Zacharias2013}, using only astrometric data that meet several 
high-quality criteria encoded in the UCAC-4 flags ($\sim 80\%$ of the stars), 
in the same manner as in \citet{Anders2014}. 

\section{Analysis}\label{ana}

\begin{table}
	\begin{tabular}{ l c }
	\hline \hline
	  Sample criterion & Stars\\
	 \hline
	  CoRoT-APOGEE stars & 690\\
	  with good ASPCAP results & 678 \\
	  and good seismic results & 664 \\
	  and $|\log g_{\mathrm{ASPCAP}}^{\mathrm{calib}} - \log g_{\mathrm{seismo}}|<0.5$ dex & 617 \\
	 \hline
	  Converged stellar PARAMeters and distances & 606 \\
	  \quad LRa01 & 281 \\
	  \quad LRc01 & 325 \\
	  and reliable UCAC-4 proper motions (OK flag) & 504 \\
	  and good orbits ($\sigma(v_T)<50$ km/s) & 234 \\
	\hline
\end{tabular}
	\caption{Summary of the number of CoRoT-APOGEE stars satisfying 
different quality criteria.}
\label{summary}
\end{table}

\subsection{Masses, radii, and ages}
To derive primary stellar parameters such as mass, luminosity, 
radius, and age, we used the Bayesian parameter estimation code 
PARAM\footnote{\url{http://stev.oapd.inaf.it/cgi-bin/param}} 
\citep{daSilva2006} with the recent improvements presented in 
\citet{Rodrigues2014}. The code uses standard grid-based modelling (see 
\citealt{Chaplin2013} and references therein for an overview) to estimate 
stellar properties by comparison with theoretical stellar models, in our case 
the PARSEC isochrone models \citep{Bressan2012}. 

When computing the desired stellar parameters, PARAM naturally accounts for 
the statistical uncertainties in the input parameters \{$\Delta\nu, 
\nu_{\mathrm{max}}, T_{\mathrm{eff}}$, [Z/H]\}, and transforms them 
into the posterior probability distribution in stellar model 
space. We therefore denote uncertainties that are reflected in the shape of stellar parameter probability distribution functions (PDFs) {\it 
statistical}, because they arise from a (non-linear) propagation of 
uncertainties in the measured quantities.\footnote{Because we 
chose a particular set of isochrones, our statistical uncertainties are of course not model-independent.} 

Stellar evolution models predict a rather tight relation between mass, 
metallicity, and age for red giants, with the age spread 
increasing with decreasing mass. Therefore, an uncertainty in stellar mass of about 10\% typically results in a (statistical) age uncertainty of about 30\% (see, e.g., \citealt{Miglio2013a}). 
In addition, depending on its location in the Hertzsprung-Russell diagram, a star may have broad or multi-peaked stellar parameter PDFs; the age PDFs show a wide variety of shapes. But even in the case of very broad PDFs, the knowledge about their shape does add valuable information: in the sense that we can quantify the knowledge we lack.

\begin{figure}\centering
\includegraphics[width=0.49\textwidth]
{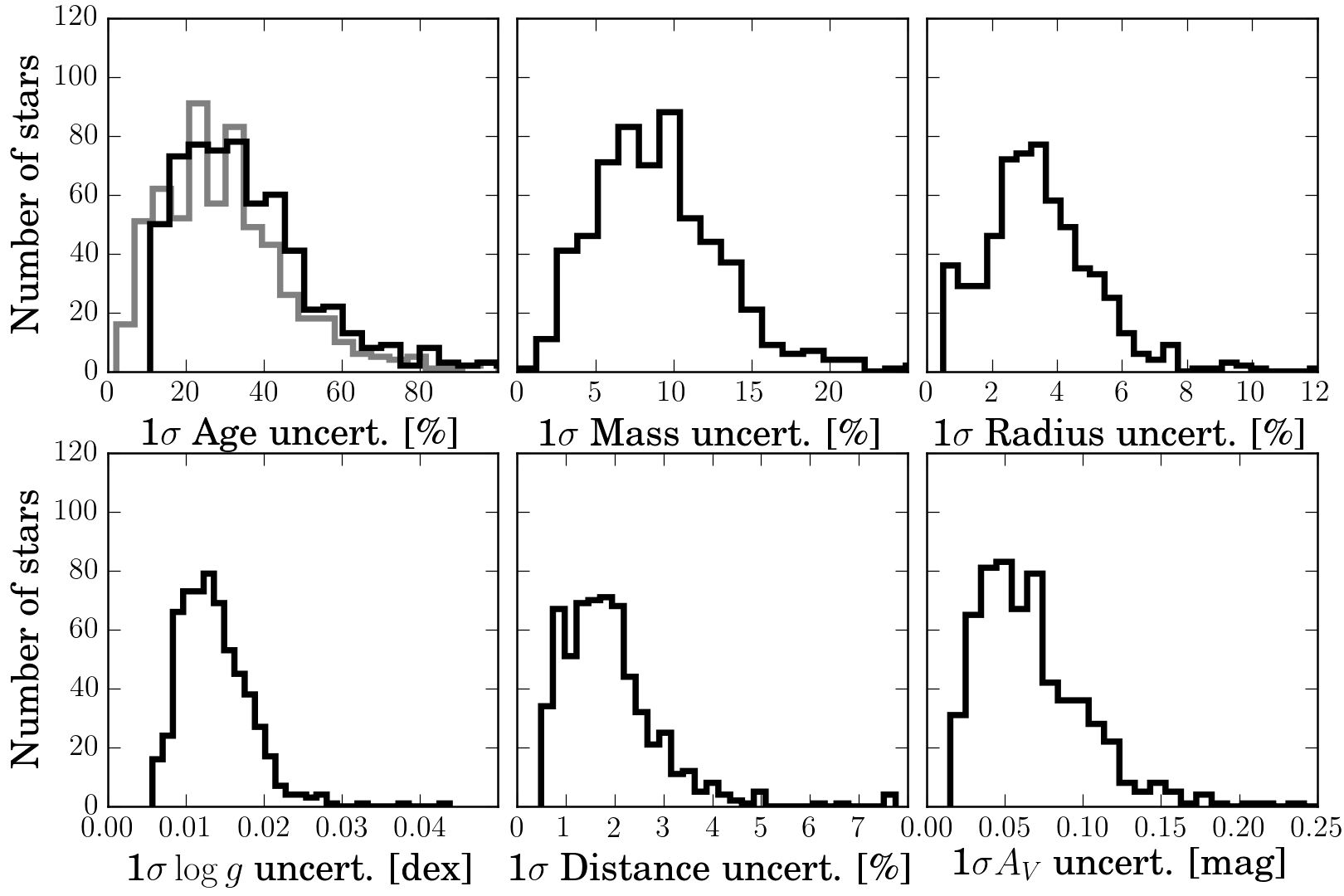}
 \caption{Distributions of the 1$\sigma$-uncertainties in stellar age, mass, radius, $\log g$, distance, and extinction for our sample. For the ages (top left panel), we show the distributions of statistical ({\it grey histogram}) and total uncertainties.}
 \label{errordists}
 \end{figure}

We therefore report the mode and 68\% or 95\% credible 
intervals of the marginalised PDF in mass, radius, age, distance, and extinction\footnote{Differently from \citet{Rodrigues2014}, we computed these statistics from the interpolated PDF in linear units, and our formal $1\sigma$ ($2\sigma$) parameter uncertainties are defined as the smallest parameter interval around the mode that contain 68\% (95\%) of the PDF.} in our catalogue.
We achieve typical statistical uncertainties of 0.015 dex in $\log g$, 4\% in 
radius, 9\% in mass\footnote{Even in the very local volume the comparison of absolute magnitude (based on {\it Hipparcos} parallax), $B-V$ colour, and [Fe/H] with stellar evolution models yields typical uncertainties in radius and mass of 6\% and 8\%, respectively \citep{AllendePrieto1999}.}, 
25\% in age, and 2\% in distance (median values; see Fig. \ref{errordists}).

As discussed in the Introduction, stellar ages are by far more uncertain than 
any other classical stellar parameter and should be used only in a statistical, relative sense. This is due to a combination of the simple propagation of the stellar mass uncertainties with systematic uncertainties (mostly related to mass loss and the mass scaling relation). The magnitude of these uncertainties and their influence on stellar age estimates are discussed in Sect. \ref{sys}. For a more detailed discussion of the systematic uncertainties involved in stellar modelling see, e.g., \citet{Noels2015}.

\subsection{Age uncertainties -- a closer look}

\subsubsection{Statistical uncertainties}

\begin{figure}\centering
\includegraphics[width=0.49\textwidth]{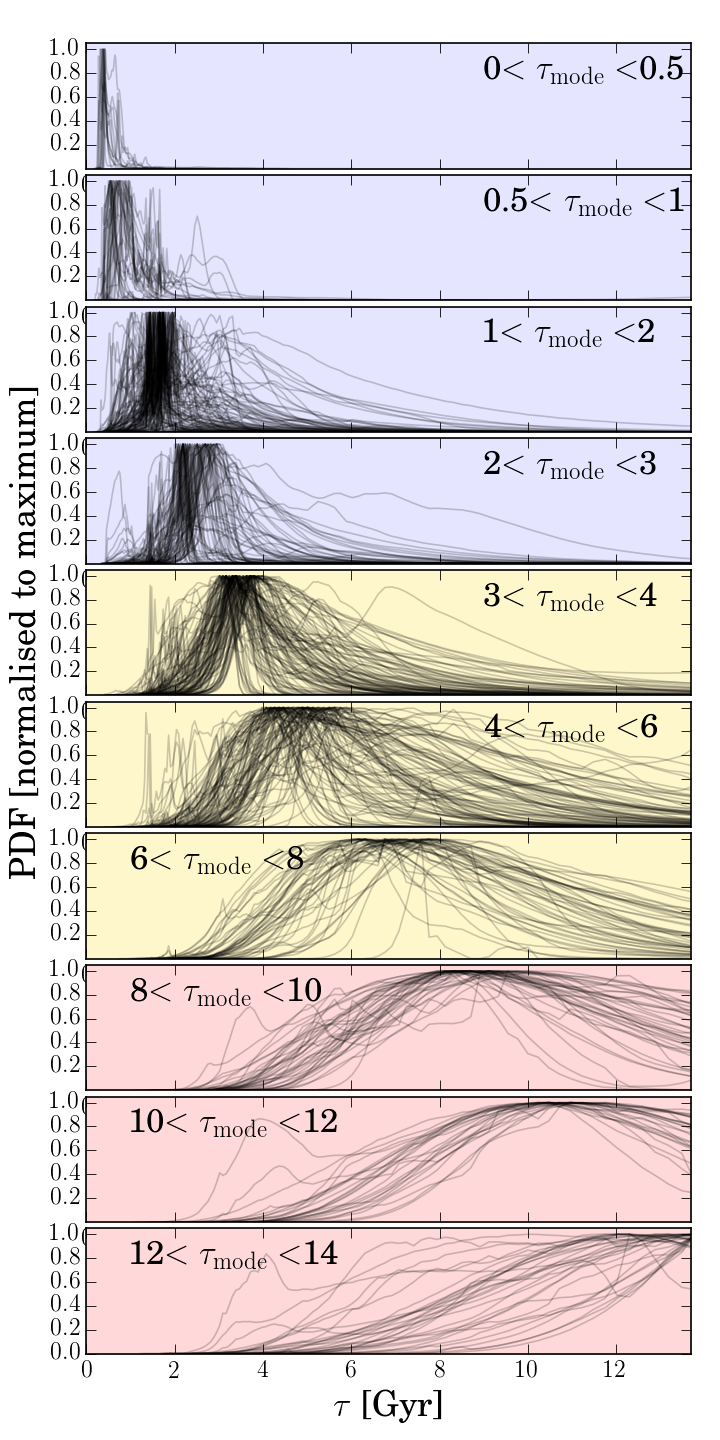}
 \caption{Diversity of the computed age PDFs: All age PDFs of the CoRoGEE sample, grouped in bins of mode age. Background colours correspond to the three age bins used in Fig. \ref{chemplanes_R}.}
 \label{agepdfs}
\end{figure}

 \begin{figure}\centering
\includegraphics[width=0.45\textwidth]
{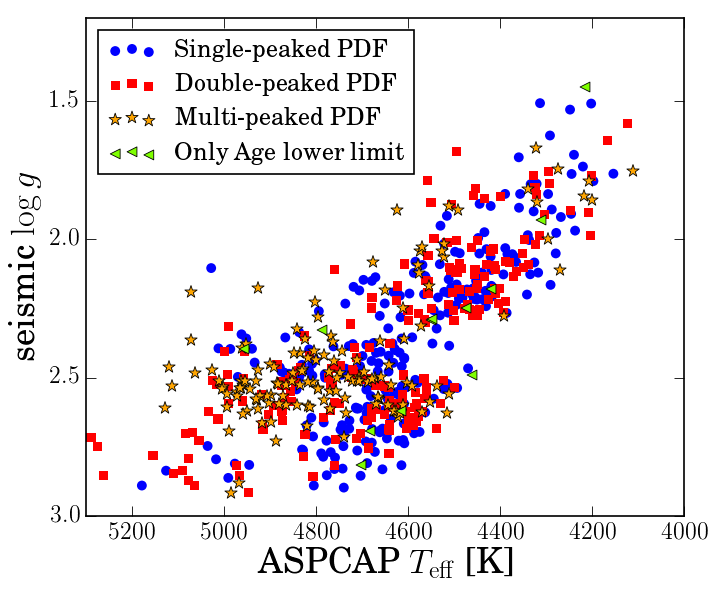}
 \caption{Seismo-spectroscopic Hertzsprung-Russell diagram of the CoRoGEE stars, with the symbols indicating the qualitative behaviour of the age PDFs, as described in the legend.}
 \label{hrd_agepeaks}
 \end{figure}

The age PDFs (which were not shown by \citealt{Rodrigues2014}) merit 
closer examination. 
Fig. \ref{agepdfs} shows the collection of all 606 age CoRoGEE PDFs, grouped in bins of mode age. Among them we find ``well-behaved'' (single-peaked) as well as more complex (double-, multi-peaked, very broad or grid-edge-affected) stellar parameter PDFs. 

It has been known for some time that isochrone-grid derived stellar age PDFs may show a great diversity (e.g., \citealt{Takeda2007}). As there is no straightforward way to classify or even quantify the behaviour of such diverse PDF shapes, the following numbers should be used with caution:

\begin{itemize}
 \item Of the 606 stars passing all quality criteria, 246 display well-behaved single-peaked age PDFs, 205 age PDFs are double-peaked, 143 have three or more peaks, and 12 do not have local extrema because the PDF increases monotonically towards the upper age limit. 
 \item Many of the multi-peaked PDFs have negligible PDF contributions from the secondary, tertiary etc. maxima, but a sizeable fraction exhibits genuinely complex function profiles.
 \item Fig. \ref{hrd_agepeaks} shows the distribution of stars classified according to the overall form of their age PDF in the Hertzsprung-Russell diagram. Multi-peaked age PDFs occur predominantly for stars with $\log g \simeq 2.4$, i.e., parameter regions that are occupied by first-ascent RGB stars as well as red-clump stars and asymptotic giant-branch (AGB) stars. The metallicity measurement does not add sufficient information to disentangle the different evolutionary stages. As noted by, e.g., \citet{Rodrigues2014}, the limiting factor is the accuracy of the effective temperatures, both in terms of models and measurements.
 \item An independent possibility of distinguishing between evolutionary phases (and thereby reducing the number of multi-peaked solutions) is offered by asteroseismology: \citet{Mosser2011} have measured mixed-mode period spacings (see also \citealt{Bedding2011}) for a fraction of the CoRoGEE targets (139 stars in LRc01, 28 stars in LRa01). This information was used to better constrain the age PDFs -- as done in \citealt{Rodrigues2014} for the APOGEE-{\it Kepler} (APOKASC) sample \citep{Pinsonneault2014}, and in \citet{Casagrande2014b, Casagrande2016} for the SAGA survey.
\end{itemize}

\subsubsection{Systematic uncertainties}\label{sys}

\begin{figure}\centering
\includegraphics[width=0.49\textwidth]
{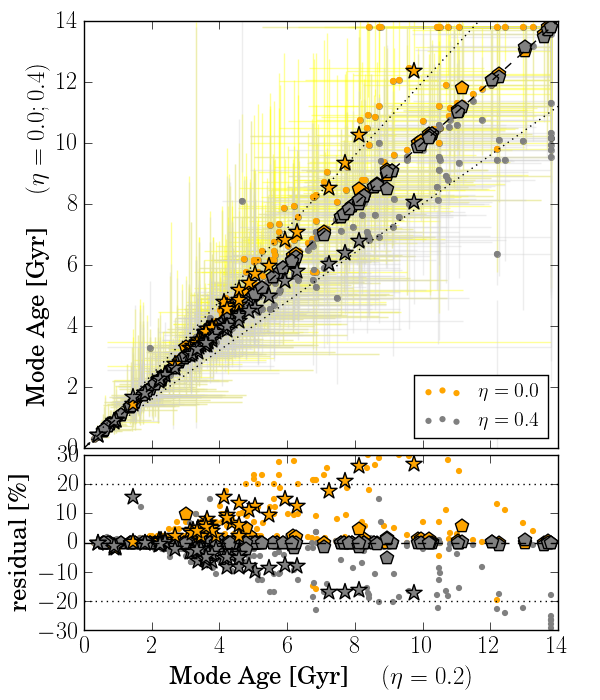}
 \caption{Effecct of non-canonical mass-loss assumptions on our derived ages. The {\it upper panel} shows the ages derived using a non-canonical mass-loss parameter $\eta_{\mathrm{Reimers}}=0.0$ (0.4) in {\it orange} ({\it grey}), while the {\it lower panel} zooms into the differences. Again, known RC stars are plotted as {\it stars}, RGB stars as {\it pentagons}.}
 \label{massloss}
 \end{figure}

For population studies of red giants, there are three main sources of systematic age uncertainties: 
\begin{enumerate}
 \item {\it The accuracy of seismic masses}: An important source of age bias comes from possible systematic errors in mass, which are likely to be small ($< 10\%$), but are very hard to quantify given that only a few objects or stars in clusters have masses known to within 10\% or better. Because hard constraints on the accuracy of the seismic masses have started to appear only very recently \citep[e.g.,][]{Miglio2016}, we refrain from a quantitative analysis in this paper. Future analyses will use a revised version of the $\Delta\nu$ scaling relation
 .
 \item {\it Mass loss}: The accuracy of age-mass relations for red giants relies on our incomplete knowledge of stellar physics.
 While a relatively simple mass-age relation is expected for RGB stars, the situation for RC or early AGB stars is different: If these stars undergo a significant mass loss near the tip of the RGB, then the mass-age relation is not unique (for a given composition and input physics), since the mass observed at the RC or early-AGB stage may differ from the initial one \citep[for a review see, e.g.,][]{Catelan2009, Miglio2012}.\footnote{In this context, the characterisation of populations of giants benefits greatly from estimates of the period spacings of the observed gravity modes, which allows a clear distinction to be made between RGB and RC stars \citep{Bedding2011}, and early-AGB stars \citep{Montalban2013a}.}
 In the PARSEC isochrones, mass loss is included following the prescription of \citet{Reimers1975}. Fig. \ref{massloss} demonstrates the effect of varying our canonical value of the mass-loss efficiency $\eta=0.2$ to extreme values ($0$ or $0.4$, respectively). Our overall results are similar to the findings of \citet{Casagrande2016} for the SAGA sample:
 The impact of mass-loss on the age uncertainty increases with evolutionary stage, in the sense that RGB stars (especially seismically confirmed RGB stars) are almost unaffected by changes in $\eta$, while for RC stars we can change the age by up to $\pm30\%$ in some cases.
 However, for the vast majority of our stars the age uncertainty due to mass loss is $\lesssim 20\%$. 
 \item {\it Other input physics:} It is well-known that the stellar physics input of theoretical isochrones (e.g., reaction rates, opacities, rotation, diffusion, He abundance, mass loss, or core overshooting) significantly affect the age and luminosities of the predicted stellar models at a given mass (e.g., \citealt{Miglio2015a, Noels2015}). 
 At this time, the quantitative effects of each of the adopted input 
 physics parameters on the isochrones are known in some detail through 
 asteroseismology (e.g., \citealt{Montalban2013, Broomhall2014, Lebreton2014a}). 
 However, a real calibration of stellar models through seismology has only started recently\footnote{For example, it has become possible to determine the amount of convective-core overshooting during the main-sequence phase \citep{SilvaAguirre2013, Deheuvels2015}}. 
 A detailed comparison of the available stellar models has not yet been performed, but a recent study (Miglio et al., in prep.) suggests that the age spread models computed with different stellar evolution codes for an early AGB star at solar metallicity is around 7\% for a $1M_{\odot}$ star, 11\% for a $1.5 M_{\odot}$ star, and 25\% for a $2M_{\odot}$ star. For this paper, we extrapolated these values to the full mass range, and neglected any possible dependency on metallicity.
\end{enumerate}

We can now define our total age uncertainty as the quadratic sum of the (asymmetric) formal $1\sigma$ uncertainty coming from PARAM, the uncertainty derived from the mass-loss test (Fig. \ref{massloss}), and the mass-dependent uncertainty coming from the comparison of different evolutionary codes.

\begin{figure}\centering
\includegraphics[width=0.49\textwidth]
{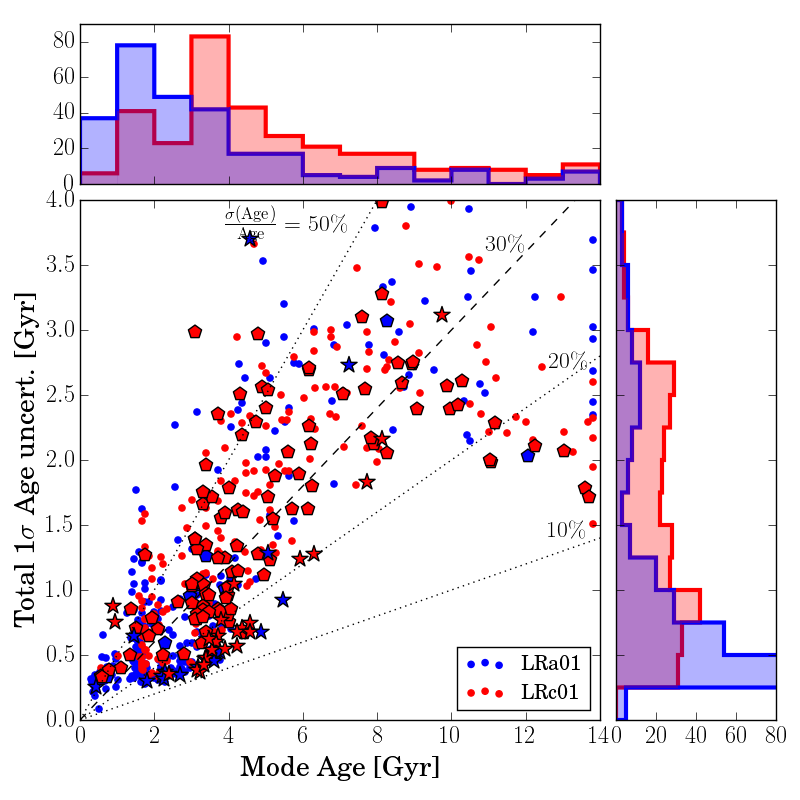}
 \caption{Total age uncertainties as a function of age (or more precisely, the mode of the age PDF), for stars in LRc01 ({\it red}) and LRa01 ({\it blue}). Known RC stars are plotted as {\it stars}, RGB stars as {\it pentagons}. The black lines indicate lines of constant fractional age uncertainties (from left to right: 50\%, 30\%, 20\%, 10\%). The histograms in the top and right panels show the distribution of ages and age uncertainties, respectively.}
 \label{ageerror}
 \end{figure}

Fig. \ref{ageerror} displays the distribution of these total $1\sigma$ age uncertainties as a function of age, colour-coded by field.
The plot shows some important features:
\begin{itemize}
 \item For stars between 4 and 10 Gyr, we observe an overall linear relation 
between age uncertainty and age. Because the finite age of the Universe (taken here as $\tau_{\mathrm{max}}=13.8$ Gyr) is included in the age prior (which is flat in $\log \tau$), the method-intrinsic age uncertainties reach a maximum at $\tau=9$ Gyr and decrease again towards greater ages.
 \item A sizeable number of stars have a PDF maximum at the age limit: 
Most of these objects can be safely assumed to be old thick-disc stars.
 \item In the younger regime, we see a complicated behaviour in the age 
uncertainty--age diagram: Some stars appear to cluster around certain age 
values. These do not correspond to the grid points of our PARSEC models 
(which is much finer: $\Delta\log(\tau{\rm[yr]})=0.01$). The observed dip in the LRc01 age distribution is not statistically significant.
 \item Although the age uncertainties are certainly non-negligible, the top 
panel of Fig. \ref{ageerror} suggests the indirect result of \citet{Miglio2013a}, who used stellar population synthesis models to conclude that the stars in LRa01 are typically younger than the LRc01 population. To make this statement more quantitative, the histograms have to be corrected for selection effects, as we discuss in Sect. \ref{disc}.
\end{itemize}

\subsubsection{Estimating age errors from simulated stars}\label{mockage}

As an additional check of our age estimates, we opted to simulate the CoRoGEE sample based on the chemodynamical model of \citet[][MCM]{Minchev2013, Minchev2014b}\footnote{The results are largely independent of the model used. However, we note that the MCM model is a thin-disc model only, and therefore does not include stars older than 11.7 Gyr.}. The final snapshot of the MCM galaxy consists of 953,206 N-body particles with age, chemical, and kinematic information. To translate these mass particles into simulated stars, \citet{Piffl2013} first used the MCM model as an input for the Galaxia code \citep{Sharma2011} in the context of a simulated RAVE survey. Here, we used the same code to simulate a CoRoGEE-like sample from the MCM galaxy. A detailed description of the chemodynamical mock is given in \citet{Anders2016a}. In the following, we briefly summarise the procedure.

We first simulated the stellar populations in the CoroT fields and calculated observed magnitudes for these mock stars using the new PanSTARRS-1 3D extinction map of \citet{Green2015} as our Galactic extinction model. In the next step, we applied the effective CoRoGEE selection function (assuming that it only depends on $H$ and $J-K_s$) by selecting stars randomly from small boxes in the colour-magnitude diagram (see Fig. 4 of \citealt{Anders2016a}). While this is certainly a simplification of the true CoRoGEE selection (see Sect. \ref{obs}), it was the only way in which our forward model could be realised. We also simulated Gaussian observational errors in the stellar parameters {$T_{\rm eff}, \Delta\nu, \nu_{\rm max},$ [Z/H]} and magnitudes, and then ran the Bayesian parameter estimation code PARAM, exactly as was done with the real data. 

 \begin{figure}\centering
\includegraphics[trim=0cm 1.1cm 0cm 0cm, clip=true, width=.49\textwidth]
{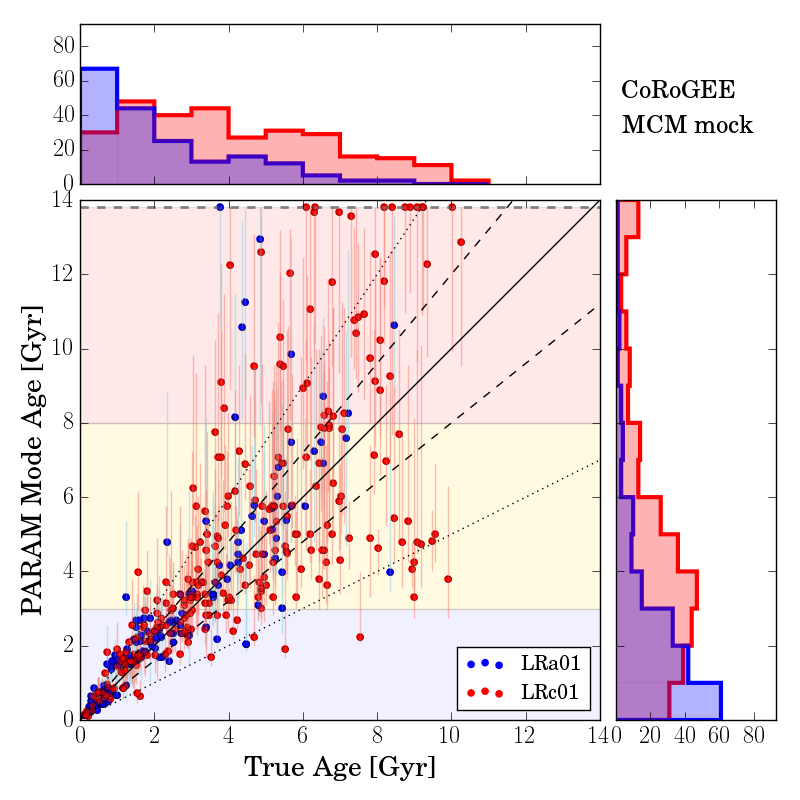}
\includegraphics[trim=0cm 0cm 0cm 3.67cm, clip=true, width=.49\textwidth]
{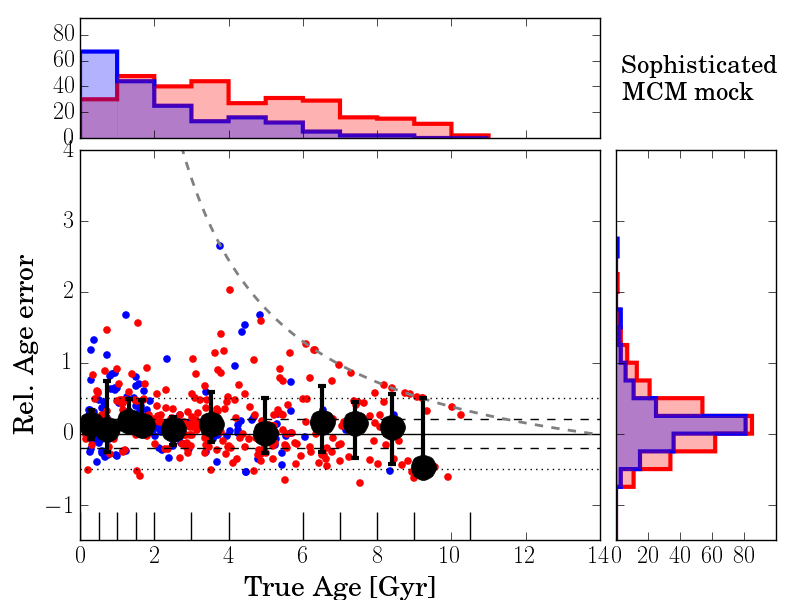}
 \caption{Estimating systematics of our age estimates using simulated stars. The scatter plot in the {\it upper panel} shows estimated PARAM ages of the CoRoGEE mock stars (and their statistical 1$\sigma$ uncertainties as error bars) vs. the true ages of the parent N-body particle. The {\it histograms} to the sides of this plot show the corresponding age distributions of the two CoRoT fields. The background colours correspond to the three age bins used in Sect. \ref{res}.
 The {\it lower plot} shows the relative age error $\frac{\tau_{\rm PARAM} - \tau_{\rm true}}{\tau_{\rm true}}$ as a function of the true age.
 The black symbols correspond to the median age error in each age bin indicated on the x-axis.
 The various lines correspond to a one-to-one relation, 20\% and 50\% deviation, and the age boundary at 13.8 Gyr.
 }
 \label{mockages}
\end{figure}

Using this simulation, we can now address the question of how well our recovered PARAM age estimates correspond to the true stellar ages given by the model: 
The upper panel of Fig. \ref{mockages} shows estimated vs. true ages, the lower panel presents the relative age error $\frac{\tau_{\rm PARAM} - \tau_{\rm true}}{\tau_{\rm true}}$ as a function of the true age.
The black symbols correspond to the median age error in each age bin indicated on the $x$-axis, demonstrating that our method tends to systematically overestimate the true ages by around $10-15$\%, with the scatter increasing towards greater ages. A small systematic shift is expected, as the Galaxia input isochrones (Padova; \citealt{Marigo2008}) are slightly different from those used by PARAM (PARSEC; \citealt{Bressan2012}). 
The histograms to the sides of the top plot show how the true age distributions (in the model) of the two CoRoT fields are distorted by the measurement procedures.

As is clear from Fig. \ref{mockages} and as shown in the previous section, our derived age estimates should be treated with caution, and considered relative age indicators rather than unbiased absolute age estimates. Therefore, in this paper we only use the age information to separate our stars into  three wide age bins: Stars with derived PARAM ages younger than 3 Gyr (``young''), stars with PARAM ages between 3 and 8 Gyr (``intermediate''), and stars measured to be older than 8 Gyr (``old''). The typical forms of age PDFs for stars in these three bins are shown in Fig. \ref{agepdfs} (coloured panels). The same coloured regions in the top panel of Fig. \ref{mockages} can be used to assess the contamination in each of the three age bins.
In summary, the simulation suggests that the contamination by old stars in the young bin and the contamination by young stars in the old bin are negligible.

\subsection{Distances and extinctions}

As in \citet{Rodrigues2014}, distances and extinctions were calculated by comparing the previously derived 
absolute magnitude with the observed magnitudes in several passbands (see Sect. \ref{phot}), assuming a single extinction curve \citep{Cardelli1989, ODonnell1994}, using the bolometric corrections of \citet{Marigo2008} and the corresponding extinction coefficients \citep{Girardi2008}. 
Because PARAM uses photometric measurements from many filters over a wide wavelength range (see Sect. \ref{phot}), our distance uncertainties are much smaller than the uncertainties expected from the distance-radius relation (as adopted in, e.g., \citealt{Miglio2013a}). 
For more details, we refer to Sect. 3 of \citet{Rodrigues2014}, and to Appendix \ref{sane}. 

We carried out comparisons with extinction estimates from the literature in Appendix \ref{ext}, finding that our precise extinction values are best matched by the spectro-photometric method developed in \citet{Schultheis2014}.


\begin{figure}\centering
 \includegraphics[width=0.5\textwidth]{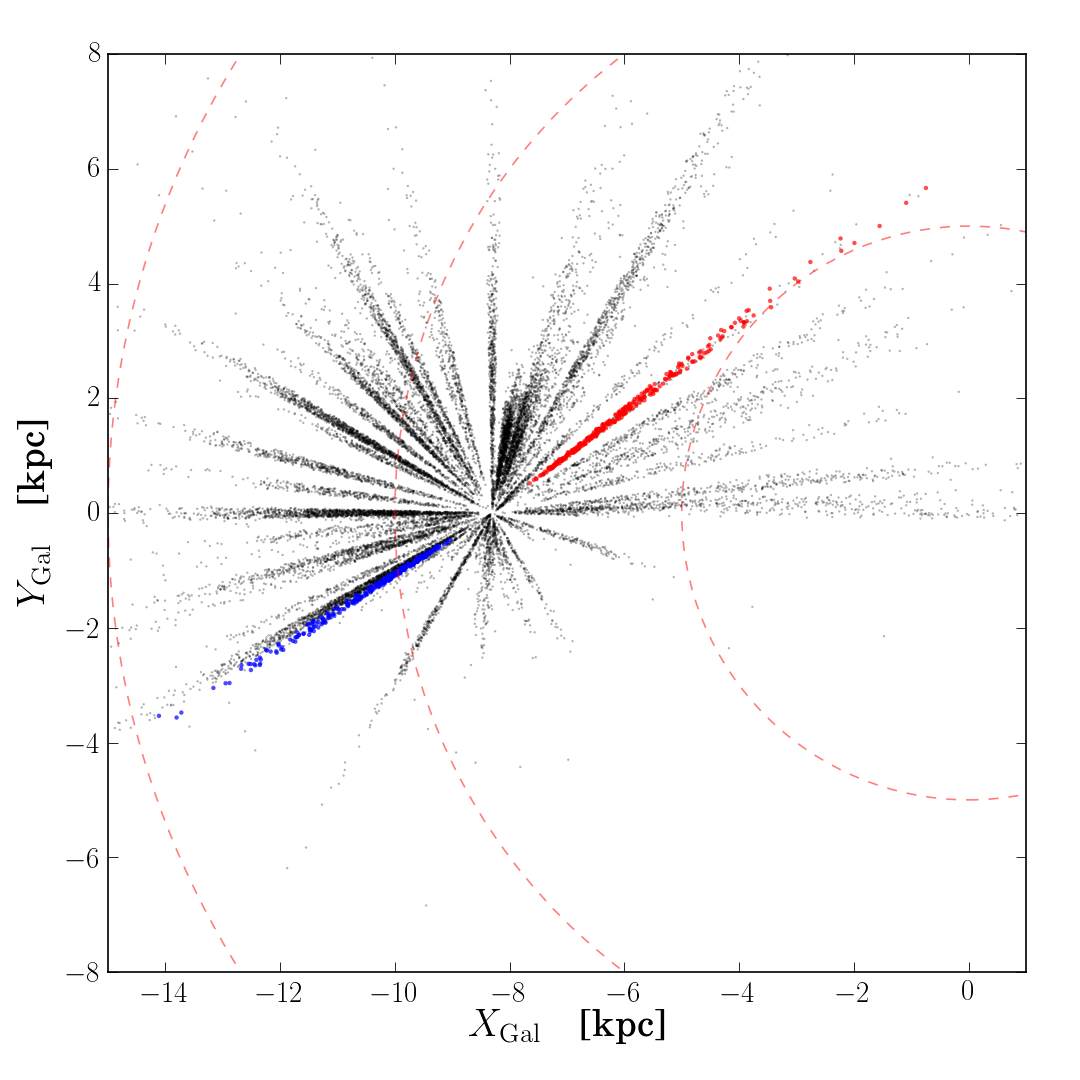}
 \includegraphics[width=0.5\textwidth]{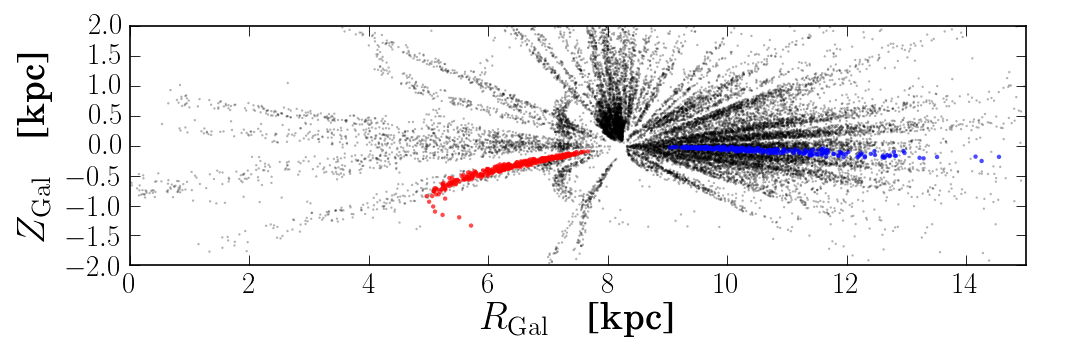}
 \caption{Location of the CoRoT-APOGEE stars in Galactocentric Cartesian 
coordinates ($X_{\mathrm{Gal}}, Y_{\mathrm{Gal}}$ -- top panel) and Cylindrical coordinates ($R_{\mathrm{Gal}}, Z_{\mathrm{Gal}}$ -- bottom). 
Blue dots correspond to LRa01 targets, red dots to LRc01 targets. The APOGEE 
DR10 high-quality giant sample \citep{Anders2014} is shown in the 
background (black dots).}
 \label{corogeegalaxy}
 \end{figure}
\subsection{Kinematics}
The 6D phase-space coordinates, along with their uncertainties are 
available for a subset of 504 stars. For this subset, orbital parameters were 
computed in the same manner as in \citet{Anders2014}.
Most of the more distant stars, however, still 
have too large proper motion uncertainties ($>50$ km/s in the tangential 
component of the space velocity, $v_T$) to be useful even for 
statistical kinematic studies, as our sample is too small to allow for 
good statistics in the presence of noisy kinematical data (see Table 
\ref{summary}). When examining the kinematical properties of our 
sample, we therefore concentrated on the most reliable parameters whenever 
possible. 

One relatively robust parameter is the guiding-centre radius of a stellar orbit, which we computed using the approximation ${R_\mathrm{guide}} = \frac{L_z}{v_c} = \frac{v_{\phi} \cdot R_{\mathrm{Gal}} 
}{v_c}$ \citep[e.g.,][]{Casagrande2011}. Here, $L_z$ denotes the angular momentum, $v_{\phi}$ the $\phi$-component of the space velocity, and $v_c\approx220~ \mathrm{km/s}$ the circular velocity at the star's position -- which for for our purposes can be assumed to be approximately constant over the Galactocentric distance range considered.

\section{The [$\alpha$/Fe]-[Fe/H]-age diagram at different Galactocentric 
distances}\label{res}

In addition to the presentation of the CoRoT-APOGEE data in the past two sections, the goal of this and following work is to study the age-abundance-kinematics relationships of the Milky Way disc outside the solar cylinder. To illustrate the value of our sample for Galactic Archaeology, in this section we study the [$\alpha$/Fe]-vs.-[Fe/H] abundance relationship with Galactocentric distance and age. 

The CoRoGEE sample  has the novel advantage of covering a wide radial 
range of the Galactic disc (4 kpc $<R_{\mathrm{Gal}}<$ 14 kpc) with red giants for which both asteroseismic and high-resolution spectroscopic data are available. 
Our final sample comprises 606 stars with converged stellar parameters and distances in the two CoRoT fields LRa01 and LRc01. However, given the extended radial and age baselines, this sample size forces us to constrain our analysis to broad bins of Galactocentric distances and ages instead of using full distribution functions. Moreover, we recall that systematic uncertainties probably affect the estimated ages presented here. Hence, we focus our analysis on larger age bins.

Following the path of \citet{Chiappini2015a}, we now examine the [Fe/H]-[$\alpha$/Fe]-age space also outside the solar neighbourhood, analysing the CoRoGEE stars for which we now also have age information.
We compare our findings to the predictions of chemical-evolution models, as well as to recent chemodynamics results. 

\subsection{Understanding [$\alpha$/Fe] vs. [Fe/H] diagrams with a chemical-evolution model}

[X/Fe] vs. [Fe/H] diagrams, and in particular the [$\alpha$/Fe] vs. [Fe/H] diagram, are widely used diagnostic tools to constrain the enrichment history of stellar populations. 
High-resolution spectroscopic data reveal two clearly-separated disc components (thin and thick) in the [$\alpha$/Fe] vs. [Fe/H] diagram, which follow their own age-metallicity relations (e.g., \citealt{Gratton1996, Fuhrmann1998, Ramirez2007, Anders2014}).
The valley between the two sequences in this diagram can hardly be attributed to simple sample selection effects \citep{Anders2014, Nidever2014} and is probably a real characteristic of the Galactic disc\footnote{However, see \citet{Bovy2012d} for a different explanation.}, as we discuss below.

\begin{figure}\centering
\includegraphics[width=.49\textwidth]{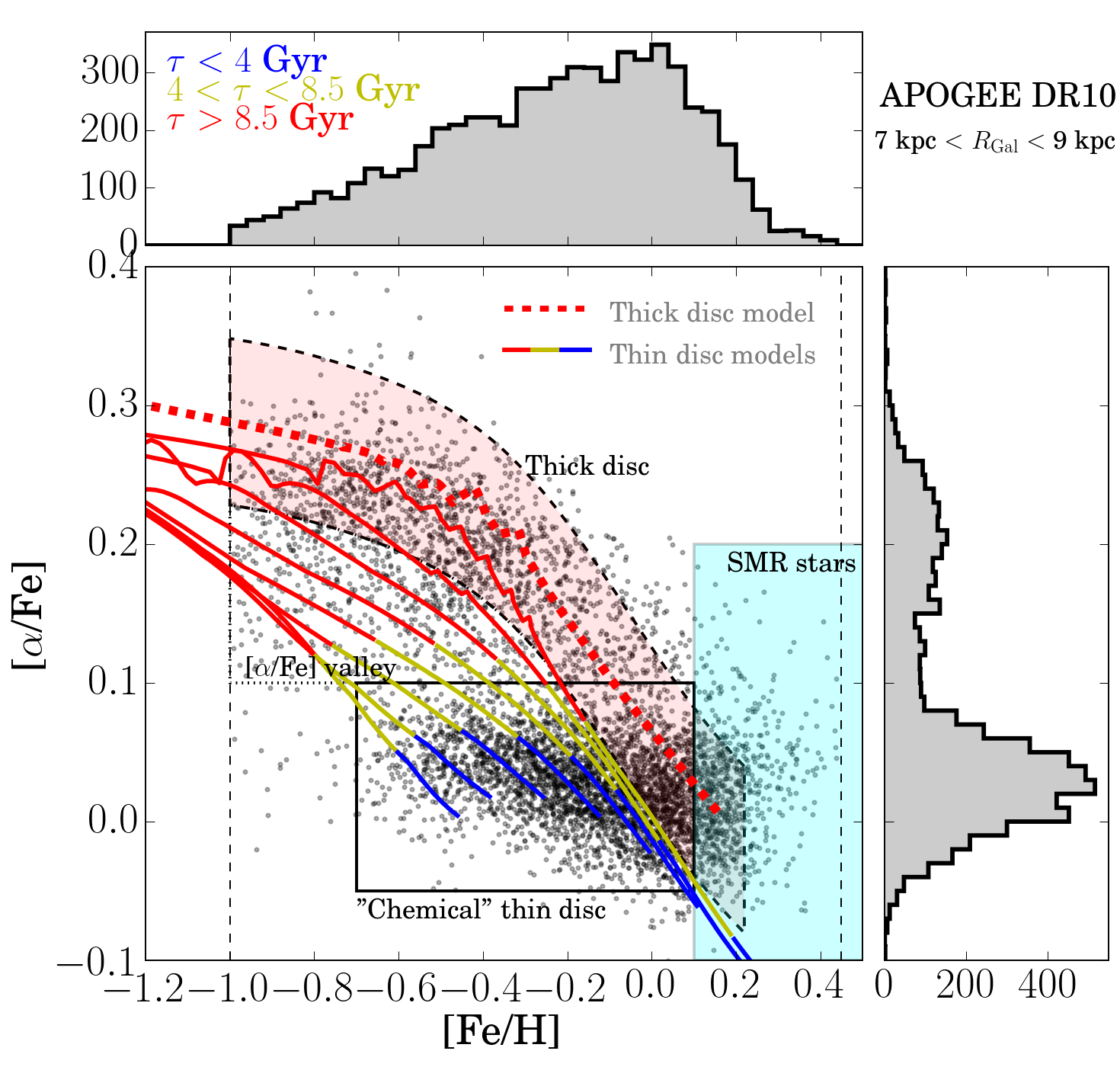} 
 \caption{The [$\alpha$/Fe] vs. [Fe/H] diagram of the APOGEE DR10 
high-quality giant sample (\citealt{Anders2014}; grey dots) in the range 7 kpc $<R_{\rm Gal}<$ 9 kpc. Overplotted with colours are the thin- and thick-disc chemical-evolution models of \citet{Chiappini2009}: The solid lines correspond to the chemical 
tracks of the thin disc at different Galactocentric annuli (from left to right: 
18 kpc, 16 kpc, 14 kpc, 12 kpc, 10 kpc, 8 kpc, 6 kpc, 4kpc). The colours 
indicate the age (or look-back time), as indicated in the top panel. 
The dashed line represents a thick-disc model (for $R_{\mathrm{Gal}} = 6$ kpc - the Galactocentric dependency for the thick-disc models computed in \citet{Chiappini2009} are minor - see text for more details). }
 \label{chemplane_crismodels}
 \end{figure}

As a starting point, in Fig.~\ref{chemplane_crismodels} we compare the bulk of APOGEE DR10 data analysed in \citet{Anders2014} with the predictions of the set of Galactic chemical-evolution models of \citet{Chiappini2009}. The figure shows the location of \citet{Anders2014} high-quality disc sample in the [$\alpha$/Fe] vs. [Fe/H] diagram, together with the histograms of these parameters. Overplotted are the chemical-evolution tracks of \citet{Chiappini2009} for various bins in Galactocentric distance, colour-coded by age. 

The thin-disc models shown in Fig.~\ref{chemplane_crismodels} were obtained by varying the accretion timescale onto the disc, assuming it to be shorter in the inner regions and longer in the outer parts (typical for MW chemical-evolution models with inside-out formation; e.g. \citealt{Chiappini1997, Chiappini2001, Hou2000}). For this reason, the thin disc at the solar vicinity formed on a longer timescale than the thick disc, and towards the inner disc regions the infall timescales of both components approach each other (but there is still a difference in the star-formation efficiency). This explains why the thin-disc model curve at 4 kpc is close to the thick disc curve (see Fig. \ref{chemplanes_R} in the next section), but reaches a lower [Fe/H] value. The details of the thin-disc model can be found in \citet*[][Sect. 3]{Minchev2013}.

From a pure chemical-evolution point of view, the thick disc can be modelled as a separate Galactic component with high star-formation efficiency and a short infall timescale. Such a model naturally predicts a population of mostly old [$\alpha$/Fe]-enhanced stars with a metallicity distribution peaking around $-$0.5 dex (e.g., \citealt{Soubiran1999}) and explains some of the abundance patterns observed in high-resolution solar-vicinity samples that are classified as thick-disc-like \citep{Chiappini2009}. When building a chemical-evolution thick-disc model of this type, one has considerable freedom in the choice of parameters because tight observational constraints are still lacking. As an example, for the thick disc models one can assume that its formation is completed within 2-3 Gyr (in order to obtain a population that is mostly older than 10 Gyr), but there is no tight constraint on the tail of the age distribution. While thin-disc models have to reproduce the chemical-abundance patterns at the present time in the local interstellar medium, the final metallicity and abundance pattern for the thick disc is still under debate (solar or super-solar depending on how this component is defined in the different datasets).  
Therefore, the thick-disc curve illustrated by the dashed line in Fig.~\ref{chemplane_crismodels} could be easily extended to higher metallicities, whereas the same is not true for the thin-disc curves (especially for the one at the solar-vicinity position). 

Figure~\ref{chemplane_crismodels} shows that these chemical models broadly agree with the two main features of the Galactic disc [$\alpha$/Fe] vs. [Fe/H] diagram: the location of the bulk of thin-disc stars at $[\alpha$/Fe$] < 0.1$  and [Fe/H] $> -$0.8 (rectangular box labelled ``chemical'' thin disc in the figure), and the position of the stars following a thick-disc track (marked by the ros\'{e}-shaded region and the thick red dashed line). Within the framework of these models, the thin-disc sequence can be explained as a mixture of relatively young (age $\lesssim5$ Gyr) stars, originating from different birth regions within the Galactic disc that have had different enrichment histories. In contrast, for the thick disc the metallicity distribution peaks at $\sim-0.5$ \citep[e.g.][]{Rocha-Pinto1996, Kotoneva2002, Nordstroem2004, Holmberg2007}, and a large number of stars is expected at high [$\alpha$/Fe] ratios and metallcities below $\sim -$0.2.  Because of the co-existence of thick and thin disc in this diagram, a gap or dip in the  [$\alpha$/Fe] vs. [Fe/H] diagram should thus be naturally produced.



Of course, the exact absolute position of the tracks with respect to the data depends not only on the calibration zeropoint of the APOGEE abundances\footnote{As an example, from SDSS DR10 to DR12, there has been a shift of $\sim0.1$ dex in the calibrated metallicities \citep{Holtzman2015, Martig2015}, and further improvements might affect the metallicity scale at the same level. A +0.05 dex shift in [$\alpha$/Fe] is also observed when moving from DR10 to DR12. As the same shift is observed between a Gaia-ESO Survey sample and the DR12 values, we opted to retain the DR10 values for the comparison in Fig. \ref{chemplane_crismodels}.}, but also on the choice of stellar yields, IMF, and star-formation efficiency.
As shown in \citet{Chiappini2009}, these models provide a good description of the observed shifts of several abundance ratios as a function of metallicity for the solar radial bin, once the thick  and thin discs are defined via kinematics (as in \citealt{Bensby2003}).

The reason we present a comparison with a model computed before the data in Fig. \ref{chemplane_crismodels} were available is to illustrate how the predictions of a pure chemical-evolution model that was in agreement with chemical abundances (among other observables) in the local volume performs when compared to the new samples of stars now covering larger portions of the disc. Clearly, one of our near-term goals is to further explore the parameter space (especially new constraints on the stellar yields and their metallicity dependency) of these models and identify those that best fit the new observational constraints. However, the main challenges to the interpretation of discrete thin and thick discs (as modelled in \citealt{Chiappini1997, Chiappini2009}) are on the one hand the existence of so-called super-metal-rich (SMR) stars \citep{Grenon1972,Trevisan2011} in the solar neighbourhood \footnote{SMR stars are defined as stars whose metal abundance exceeds the metallicity of the local present-day interstellar medium. This value is dependent on Galactocentric distance and is constrained by the present-day abundance gradient in the interstellar medium. For the solar vicinity, SMR stars are found in the region illustrated by the blue rectangular box in Fig.~\ref{chemplane_crismodels}.}, and on the other hand the fact that not all thin-disc stars with metallicities below ~$-$0.2 can be explained as high-eccentricity intruders from outer regions (\citealt{Anders2014}). 

The existence of SMR stars is commonly attributed to a significant radial mixing of stellar populations within the Galactic disc \citep[e.g.,][]{Grenon1989, Grenon1999, Chiappini2009, Kordopatis2015}. In agreement with previous studies, \citet{Kordopatis2015} conclude that SMR stars in the solar neighbourhood must have migrated from far inside the solar annulus. Recently, \citet{Schoenrich2009, Brunetti2011, Minchev2013, Minchev2014} and \citet{Kubryk2015, Kubryk2015a} have argued that chemical-evolution models for the Milky Way cannot be viewed independently of its dynamical evolution, and found different prescriptions for the merging of these two aspects of Galactic evolution. In the next subsection we separate the [$\alpha$/Fe]-[Fe/H] diagram into bins of age and Galactocentric distance and compare our data to a chemical-evolution model. This is useful because the latter form the backbone of many recent chemo-dynamical approaches. 

\subsubsection{Binning the data in Galactocentric distance and age}\label{bins}

While the division of the massive APOGEE dataset into various Galactic zones has been the subject of previous investigations \citep{Anders2014, Hayden2014, Nidever2014, Hayden2015}, we now can make use of the unique seismic information from CoRoT to show, for the first time, [$\alpha$/Fe]-[Fe/H]-age diagrams, outside the {\it Hipparcos} volume, in several Galactocentric bins. 

\begin{figure*}\centering
 \includegraphics[width=.99\textwidth]{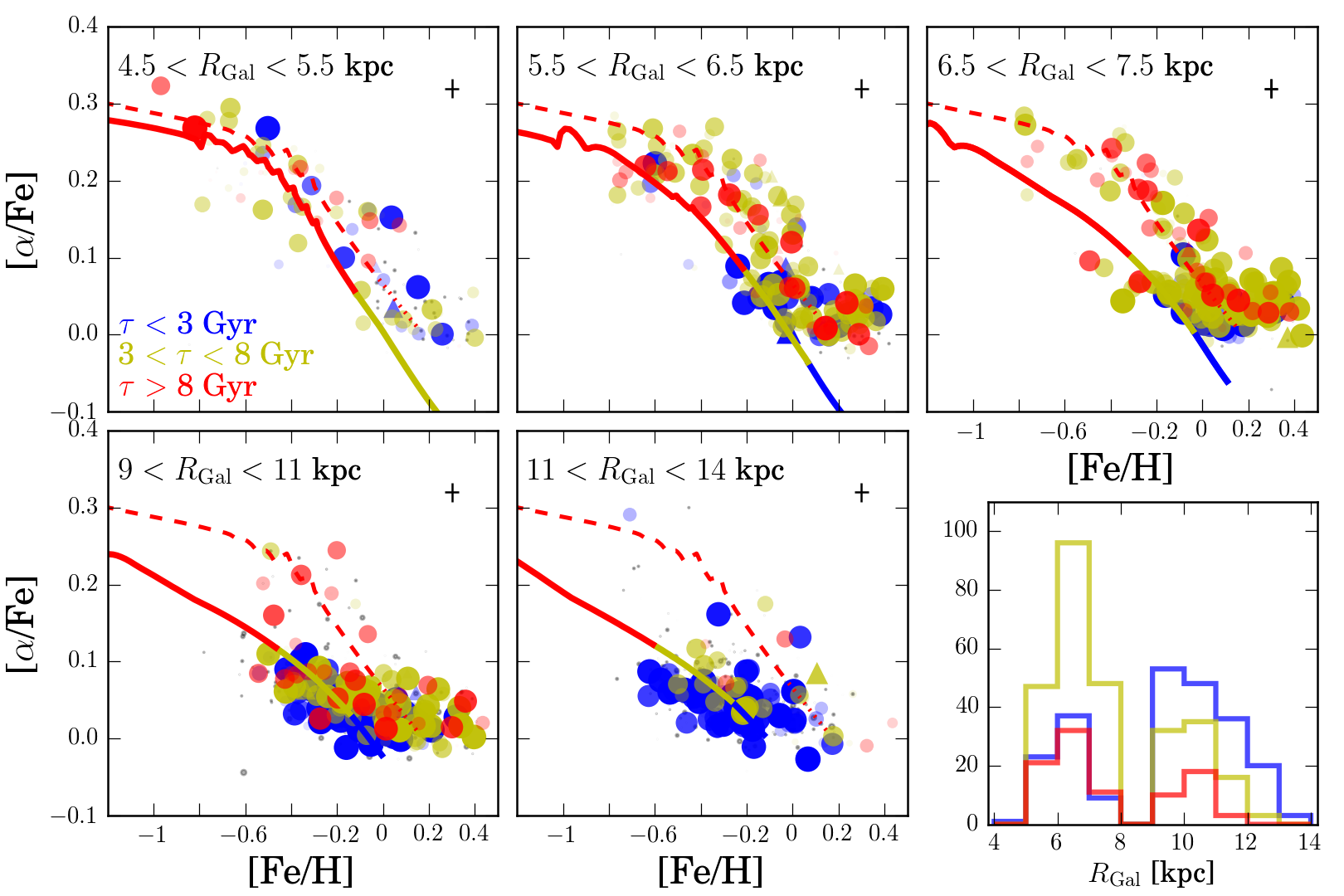}
 \caption{The [$\alpha$/Fe]-vs.-[Fe/H] chemical plane for five different bins in Galactocentric distance $R_{\mathrm{Gal}}$. The {\it colour} represents our stellar age estimates, as indicated in the first panel: blue indicates stars younger than 3 Gyr, red stars older than 8 Gyr, and yellow intermediate ages.  The {\it point size and transparency} of each data point encode the age uncertainty, i.e., a smaller and more transparent symbol corresponds to a lower probability to belong to the particular age bin. The few {\it triangles} correspond to stars whose measured radial velocity scatter is greater than 800 m/s and which could be binaries. In the background of each panel, stars from the APOGEE DR12 main sample observed in similar Galactic regions are plotted as {\it grey dots} for comparison. The {\it error bar} in the upper right corner of each panel represents the typical (internal) uncertainty of the chemical abundances. The {\it solid lines} correspond to the thin-disc chemical-evolution model of \citet{Chiappini2009} for different Galactocentric distances, and the {\it dashed lines} correspond to a thick-disc model at $R_{\mathrm{Gal}}=6$ kpc. The lower right panel displays the overall $R_{\mathrm{Gal}}$ distributions of our sample split into the three age bins.}
 \label{chemplanes_R}
 \end{figure*}

Figure \ref{chemplanes_R} presents one of the main results of this paper: the [$\alpha$/Fe] vs. [Fe/H] diagram for the CoRoGEE sample, split into five bins of Galactocentric distance, as indicated in each panel. 
 
 \begin{figure*}\centering
 \includegraphics[width=.99\textwidth]
 {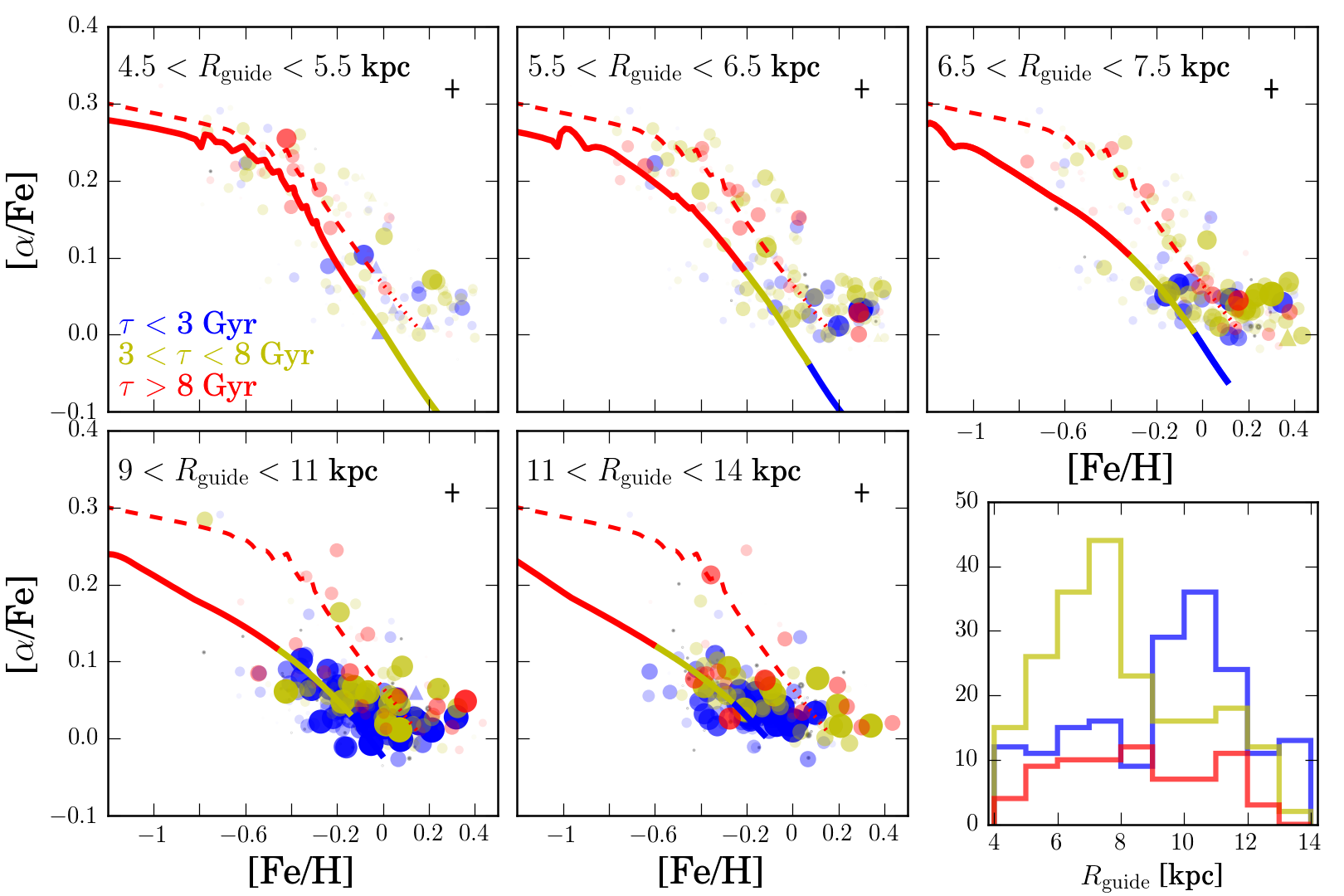}
 \caption{Same format as Fig.\ref{chemplanes_R}, only that the data are now binned in guiding-center radius, $R_{\mathrm{guide}}$, rather than Galactocentric distance, to mitigate the effect of stellar mixing by ``blurring''. Again, the colour-code represents the age, while the size and transparency now encode the uncertainty in stellar age {\it and} guiding-center radius. (If a star has a highly uncertain guiding radius -- i.e. an $R_{\mathrm{guide}}$ PDF which extends over multiple $R$ bins-- it will appear as a faint dot in multiple panels of this Figure.)}
 \label{chemplanes_Rc}
 \end{figure*}

As in Fig. \ref{chemplane_crismodels}, we include in Fig. \ref{chemplanes_R} the [$\alpha$/Fe] vs. [Fe/H] model tracks of \citet{Chiappini2009}. 
Fig.\ref{chemplanes_Rc} has the same format as Fig.\ref{chemplanes_R}, only that the data are now binned in guiding-center radius $R_{\mathrm{guide}}$ instead of Galactocentric distance, to mitigate the effect of stellar mixing by ``blurring'' \citep{Schoenrich2009}. In this plot, the size and transparency of the symbols encode the uncertainty in both stellar age {\it and} guiding-center radius, because both quantities may have considerable uncertainties. 
The interpretation of these figures is difficult because of the low statistics and the noise arising from proper motion uncertainties and radial migration. We analyse the two figures simultaneously below. 

The main results we derive from these figures are:
\begin{itemize}
\item The shift of the peak of the thin disc's metallicity distribution function from higher to lower metallicities as one moves towards larger Galactocentric distances \citep{Anders2014, Hayden2015} is accompanied by a dominance of younger ages towards the outermost radial bins. However, the exact relative number of young and old stars in each radius bin can be slightly biased as a consequence of the detectability of oscillations: younger stars are on average more luminous, and therefore exhibit larger oscillation amplitudes that are easier to detect at large distances.
\item While the inner Galaxy is dominated by stars with thick-disc-like chemistry (elevated [$\alpha$/Fe] ratios) with a large number of old stars (but see below), almost no high-[$\alpha$/Fe]-old stars are found in the outermost radial bin. This result is believed to be a manifestation of the shorter scale length of the thick disc with respect to the thin disc \citep{Bensby2011, Bovy2012d, Cheng2012}.  
\item A greater number of young-[alpha/Fe]-rich stars is seen in the two innermost bins \citep{Chiappini2015a}. These stars are not only in strong disagreement with the predictions of chemical-evolution models, but are also impossible to explain by radial migration. For a discussion of the origin of these stars see \citet{Chiappini2015a,Jofre2016,Yong2016}.
\item Surprisingly, the thin-disc chemical-evolution model adopted here provides a fairly good description of the main abundance ratio trends shown in the figures (especially in the outer parts of the Galaxy), both in terms of the abundance trends and in terms of expected dominant age. In particular, when guiding radii are used instead of the current Galactocentric distances, the agreement with the models is improved (see the 5-6 kpc and 6-7 kpc $R_{\mathrm{guide}}$ bins).  However, there is a clear disagreement above solar metallicity in all panels. Although part of the discrepancy might be attributed to uncertainties related to stellar yields\footnote{Currently there are several uncertainties affecting the stellar yields of the different $\alpha$-elements. For core-collapse supernovae, few models were computed for metallicities above solar; moreover, most supernovae models tend to underestimate the $^{24}$Mg yields. Other elements, such as Ca, Si, and S, can have some contribution of SNIa as well. Even more importantly, the Galactic SNIa rate is still very uncertain (e.g., \citealt{Matteucci1999, Mannucci2006}). Although the thin disc model presented here reproduces the present SNIa rate at the solar vicinity well, overestimated SNIa rates at earlier times and/or at other Galactocentric distances cannot be excluded. In the inside-out thin-disc formation model, one of the assumptions is that the star-formation efficiency increases towards the inner regions. This feature was mainly constrained by the abundance gradients at present time. However, if the SNIa rate is overestimated, one would require lower star-formation efficiencies to reach the same final metallicity. The abundance ratios at the different Galactocentric distances can further constrain these models, because a larger star-formation efficiency would also predict larger [$\alpha$/Fe] ratios at larger metallicities. It is thus possible that, by exploring the parameter space of stellar yields and SNIa rates, one can obtain a better fit to the data shown in the Figure, but this is beyond the scope of the present work.}, it is tempting to interpret this result as a sign of radial migration, at least for the old and intermediate-age stars (see MCM13, Fig. 8).
\item
Interestingly, in each bin, stars with high [$\alpha$/Fe] abundances, regardless of their age, show a tendency to lie close to the thick-disc curve. As the same thick-disc curve is shown in all panels, this result agrees with the relative constancy of the ``high-[$\alpha$/Fe] sequence'' discussed in \citet{Nidever2014}. It is clear from this comparison that these stars can either be explained as being part of the thick disc, or as migrators coming from the inner radii (the thick disc curve is similar to the that for $R_{\mathrm{Gal}} =$ 4 kpc, except for its higher star-formation efficiency, which leads to the appearance of [$\alpha$/Fe]-enhanced stars at higher metallicities). These oldest metal-rich, [$\alpha$/Fe]-enhanced stars also resemble Galactic bulge stars in chemistry, so that radial migration from the bulge cannot be excluded as one possible interpretation.
\item
SMR stars are present even in the two outermost $R_{\mathrm{Gal}}/R_{\mathrm{guide}}$ bins studied here \citep{Anders2014}; they comprise stars of all ages, in agreement with what was found by \citet{Trevisan2011} for solar-vicinity SMR stars. As explained previously, the end of the thin-disc curves is constrained by the present abundance gradient, which amounts to around $-0.07$dex/kpc for Fe (e.g., \cite{Anders2014}) and references therein). While the excess of SMR stars is not a problem in the inner bins (where the thin-disc curve extends to higher metallicities), it demonstrates a clear discrepancy for the two outermost bins analysed here. From the comparison with the models it is clear that the chemistry of these SMR stars is compatible either with the thick-disc curve or with the thin disc at $R_{\mathrm{Gal}} =$ 4 kpc. We note, however, that the  $R_{\mathrm{Gal}} =$ 4 kpc  curve predicts intermediate ages for stars above metallicities $\sim -$0.02, while there are clearly older SMR stars in all panels. This is an indication that these stars indeed migrated from $R_{\mathrm{Gal}} <$ 4 kpc. Unfortunately, the form of the present-day abundance gradients in the innermost regions of the Galactic disc is still unknown (see \citealt{Stasinska2012} for a discussion) -- a constraint that would shed more light on the origin of these stars.
\end{itemize}

\subsection{Comparison with a chemo-dynamical model}\label{disc}

As first shown in \cite*{Minchev2013}, when radial migration is taken into account in a chemodynamical model of the thin disc, the oldest stars in the simulation have properties similar to what we commonly identify as the thick disc (this result was later confirmed by \citealt{Kubryk2015}\footnote{In this case, differently from \cite{Minchev2013}, the authors followed a suggestion made in \citet{Brunetti2011}:the radial migration process was approximated by a diffusion process with diffusion coefficients that varied in time and position. These were extracted from an N-body+SPH simulation of a galaxy very different from the Milky Way and implemented in a standard chemical-evolution model. The coefficients were then re-scaled to fit the local G-dwarf metallicity distribution.}). Interestingly, although it is able to reproduce several properties of ``the thick disc'', our chemodynamical model does not predict a discontinuity in the [$\alpha$/Fe] vs. [Fe/H] diagram\footnote{However, when selecting particles using the same kinematical criteria as in \citet{Bensby2003}, it was possible to recover the two sequences in the [$\alpha$/Fe] vs. [Fe/H] diagram.}. The reason for this discrepancy might be the existence of a discrete thick disc component \citep{Chiappini1997, Chiappini2009}, with its specific chemical pattern, which was not included in the MCM model. To shed more light on this problem, a proper comparison between the MCM model predictions and observations is required. Because astronomical surveys are often affected by non-trivial selection effects, the comparison of survey catalogues with a Galactic model is much easier when a mock observation of the model is created (e.g., \citealt{Binney2015}). 

\begin{figure}\centering
\includegraphics[trim=0cm 6cm 0cm 0cm, clip=true, width=.46\textwidth]{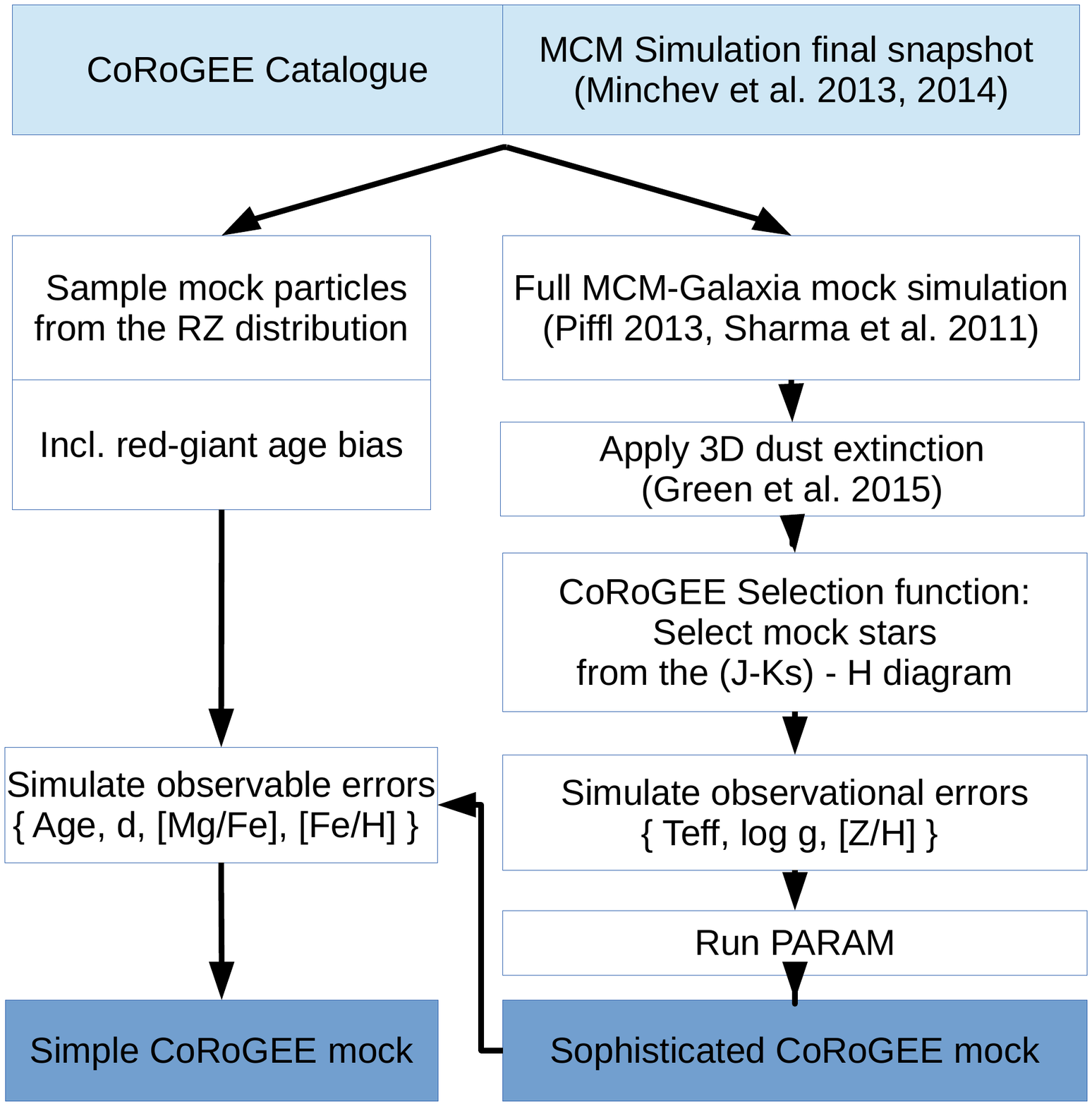}
 \caption{Scheme illustrating how the CoRoGEE mock observations were obtained from the MCM model. The steps are explained in more detail in \citet{Anders2016a}.  }
 \label{mockchart}
\end{figure}

In this section we describe our selection of a CoRoGEE-like sample from an N-body simulation, using the example of the chemodynamical N-body model analysed in \citet*[][MCM]{Minchev2013, Minchev2014b}. We have chosen two different paths to simulate the observations: 1. A ``simple'' mock in which we choose N-body particles such that we match the observed spatial distribution of our program stars (and simulating the red-giant age bias with a simple prior), and 2. A more sophisticated mock that used a modified version of the Galaxia synthetic stellar population code \citep{Sharma2011, Piffl2013}, the new PanSTARRS-1 3D extinction map of \citet{Green2015}, and a representation of the CoRoGEE selection function. The procedures leading to the two versions of mock observations are sketched in Fig. \ref{mockchart} and are explained in \citet{Anders2016a}. 
As we show below, these two versions of an MCM-CoRoGEE mock sample each have their advantages and drawbacks. In summary, while the simple mock by construction matches the space distribution of the observed sample perfectly, the sophisticated mock recovers the observed age distribution very well (see \citealt{Anders2016a}).

\begin{figure*}\centering
 \includegraphics[width=.99\textwidth] {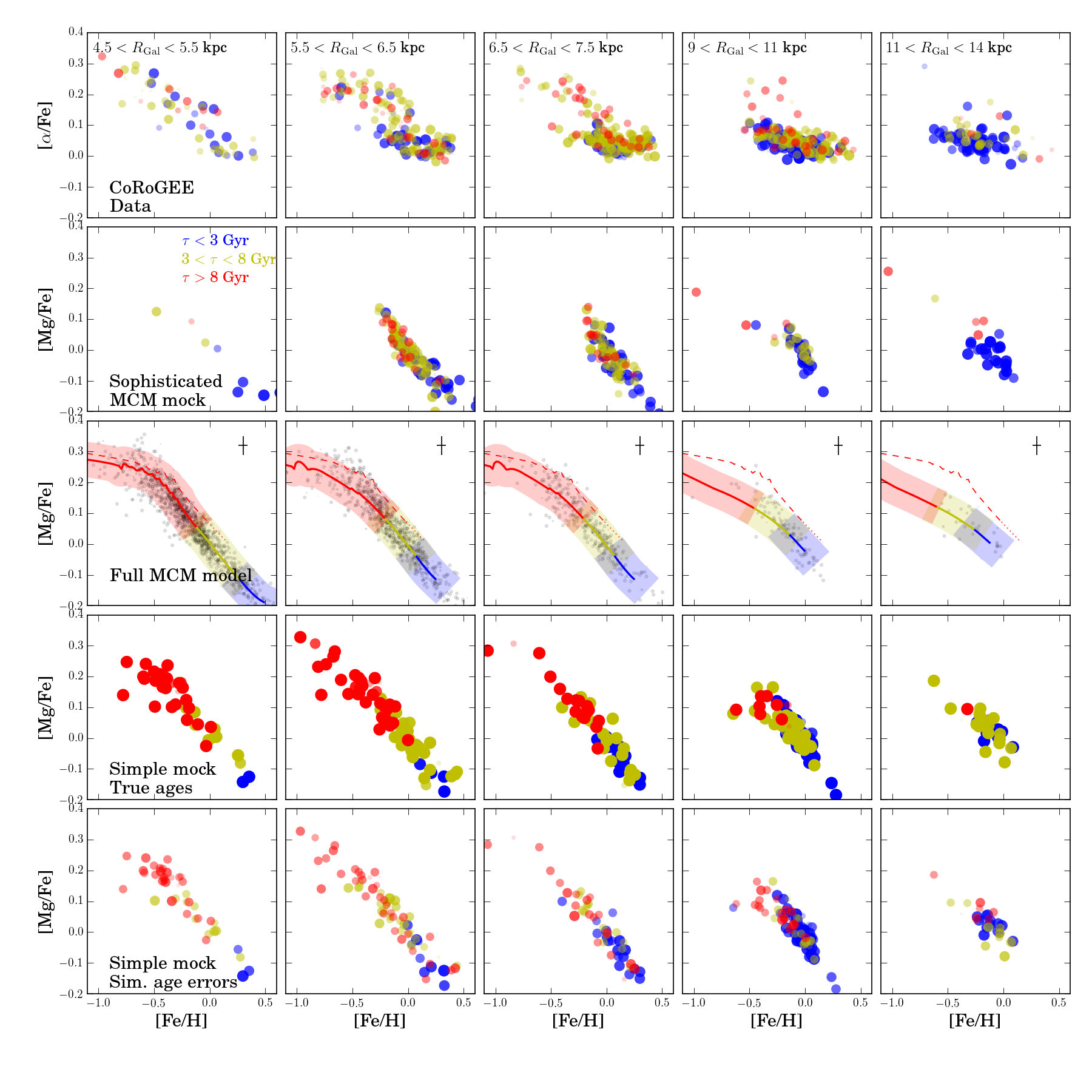}
 \caption{[$\alpha$/Fe]-vs.-[Fe/H] diagram for five different bins in Galactocentric distance $R_{\mathrm{Gal}}$.
 {\it Top row:} The CoRoGEE sample. The {\it colours} represent our stellar age estimates: blue indicates stars younger than 3 Gyr, red stands for stars older than 8 Gyr, and yellow for intermediate ages.  The {\it point size and transparency} of each data point encode the age uncertainty, i.e., a smaller and more transparent symbol corresponds to a smaller probability of belonging to the particular age bin. 
{\it Second row:} The sophisticated MCM mock sample. 
{\it Middle row:} All mock particles from the MCM N-body simulation, shown as {\it grey dots}. 
The {\it solid lines} represent the predictions of the underlying thin disc chemical-evolution model by \citet{Chiappini2009} for stars born in the corresponding Galactocentric distance bin. The colour-code represents the age; the {\it shaded regions} along the lines correspond to a $2\sigma$-confidence band, given the typical uncertainties in [Fe/H] and [$\alpha$/Fe]. The {\it dashed lines} show the chemical tracks of 
\citet{Chiappini2009} for the thick disc. The {\it error bar} in the upper right of each panel represents the typical (internal) uncertainty of the chemical abundances.
{\it Forth row:} The simple MCM mock sample, without simulated age uncertainties.
{\it Last row:} The simple MCM mock sample, with age uncertainties.
}
 \label{chemplanes_all}
 \end{figure*}

Figure \ref{chemplanes_all} shows the main result of our mock samples: each row contains the (observed or modelled) [$\alpha$/Fe] vs. [Fe/H] diagrams in the same $R_{\rm Gal}$ and age bins as in Figs. \ref{chemplanes_R} and \ref{chemplanes_Rc}, to facilitate a qualitative comparison with the data. 
We discuss the main results from Fig. \ref{chemplanes_all} below.

\begin{enumerate}
\item {\it Sophisticated mock:} \citet{Anders2016a} have shown that the observed age distributions in the two CoRoGEE fields are very well recovered by the sophisticated mock. This is also seen in the second row of Fig. \ref{chemplanes_all}: the age mix of CoRoGEE stars is better reproduced in the sophisticated mock than in the simple mock. However, the sophisticated mock obviously misses the distant and metal-poor ([Fe/H]$<-0.5$) stars that are present in the data. This indicates that our forward modelling of the sample selection is imperfect for various reasons: 1. a mismatch between the MCM-Galaxia model's starcounts with 2MASS in the CoRoT fields \citep{Anders2016a}, 2. a stronger extinction in the PanSTARRS extinction maps \citep{Schlafly2014, Green2015} compared to the CoRoGEE data (see App. \ref{ext}), 3. a more complex selection function than $S\propto S({\rm field}, H, J-K_s)$ (see Sect. \ref{spec}), and 4. stochasticity due to the small sample size. We therefore refrain from interpreting the number counts in the sophisticated CoRoGEE mock, as we did not recover the overall distributions in the abundance diagrams.
\item {\it Simple mock with true ages:} By construction (selection of mock particles from the $R_{\rm Gal}-Z_{\rm Gal}$ plane), the simple mock matches the space distribution of the CoRoGEE sample perfectly. The simple mock also matches the observed metallicity distributions much better than the sophisticated mock. The plot also demonstrates that despite the quite strong radial mixing in the MCM model, there is little age mixing in each of the [$\alpha$/Fe] vs. [Fe/H] diagrams. The age$-[\alpha$/Fe] relation of the input models is largely preserved, while the data show a significantly more complex situation. In concordance with the data, the density of the old [$\alpha$/Fe]-enhanced thin disc (i.e. the ``thick disc'' in MCM) decreases towards outer regions. However, the data suggest that the [$\alpha$/Fe]-enhanced component has a much broader age distribution than in the model. This result depends little on the functional form of the simulated age bias.
\item {\it Effect of adding age errors:} When we add realistic age errors using the PARAM results of the sophisticated mock (see Fig. \ref{mockages} and \citealt{Anders2016a}), part of the age mixing in the [$\alpha$/Fe] vs. [Fe/H] diagram can be explained by our measurement procedure. This is insufficient to explain the observed younger ages of many [$\alpha$/Fe]-enhanced stars, however. In particular, our method-intrinsic age errors cannot explain the presence of young [$\alpha$/Fe]-rich stars, while possible close-binary stellar evolution cannot explain the different abundance of these stars in the two CoRoT fields (see also \citealt{Chiappini2015a, Martig2015, Yong2016, Jofre2016}).
\item {\it SMR stars in the outer disc:} As discussed in the previous section, the metal-rich stars in the two outer bins cannot be explained with the present chemical models. Since the MCM mocks also do not produce this metal-rich intermediate-age population in the outer parts of the disc, either a much stronger radial migration than present in MCM is at work, or the thick disc star-formation history extends to greater ages (i.e., for longer than 2 Gyr). Another explanation might also be bulge stars ending up in the outer disc \citep{Barbuy1990}; these were not included in the MCM simulation.
\end{enumerate}

Our simple mock outperforms the sophisticated mock in almost all respects (except for the match with the overall age distributions). It highlights two important features in the data that are not reproduced by the MCM model: the broad observed age distribution of the [$\alpha$/Fe]-enhanced sequence in the inner Galactic disc, and that more intermediate-age SMR stars are located in the outer disc than predicted.

\section{Conclusions}\label{conclu}

In this first CoRoGEE paper, we demonstrated the usefulness of combining asteroseismic and spectroscopic data in the framework of Galactic Archaeology.
Using global asteroseismic parameters $\Delta\nu$ and $\nu_{\mathrm{max}}$ determined from CoRoT light curves, together with atmospheric stellar parameters measured by SDSS-III/APOGEE and broad-band photometry, we have calculated masses, radii, ages, distances and extinctions for more than 600 red giants distributed over a large Galactocentric distance interval. 
In this section, we briefly summarise the main results of our work.

The relative statistical uncertainties in our primary derived quantities from the Bayesian model fitting performed by the PARAM code amount to $\lesssim 2\%$ in distance, 0.08 mag in $A_V$, $\sim 4\%$ in radius, $\sim 9\%$ in mass and $\sim25\%$ in age. In agreement with previous studies, we find that the individual age probability distributions can be complex in shape, suggesting that the age information needs to be used with some care, for example, by using wide age bins.
Equally importantly, systematic uncertainties in the fundamental seismic parameters as well as in the comparison with stellar models may affect the absolute scale of our derived ages to some degree.  

We provide a number of checks (surface gravity comparison, grid-based vs. scaling relation results, extinction maps) that demonstrate the overall reliability of our analysis for the use with statistical samples in Appendix \ref{sane}.
The CoRoGEE sample enabled us to study for the first time the 
[$\alpha$/Fe]-[Fe/H]-age relation beyond the solar vicinity. We separated the sample into large bins of age, guiding-centre radius, and Galactocentric distance, to study stellar populations in the [$\alpha$/Fe] vs. [Fe/H] diagram. Even with this small sample and the sizeable systematic and statistical uncertainties attached to our age estimates, we can place reliable constraints on the chemical evolution of the Milky Way stellar disc:
\begin{enumerate}
   \item In accordance with previous work, we find strong signatures of inside-out formation of the Galactic disc.
   \item When we compared our results to a multi-zone chemical-evolution model that treats the thin and thick disc separately, we found that the thin-disc models generally provide a good description of the main abundance-age trends, with the exception of the flat [$\alpha$/Fe] trend at high metallicity. The results improved when the stellar guiding-centre radius was used instead of the current Galactocentric distance.
   \item In agreement with previous studies, we find that these pure chemical-evolution models fail to reproduce several important features seen in the data, such as the existence of SMR stars ([Fe/H]$>0.2$) in the solar neighbourhood and beyond, the exact shape of the [$\alpha$/Fe]-[Fe/H] distribution (in particular in the inner regions of the disc), and
   the existence of [$\alpha$/Fe]-rich young stars.
  \item When we compared our results with the predictions of the chemo-dynamical model of \citet*{Minchev2013, Minchev2014}, we found that the radial mixing in the model is not efficient enough to account for the number of SMR stars in the outer disc. Either a stronger radial mixing or the inclusion of a thick disc/bulge that formed stars for more than 3 Gyr and produced SMR stars may resolve this discrepancy.
  In addition, the age distribution of the [$\alpha$/Fe]-enhanced sequence in the CoRoGEE inner-disc field is much broader than expected from a combination of radial mixing and observational errors. Evolved blue stragglers may account for part of this population \citep{Jofre2016, Yong2016}, but do not offer an explanation for the different number counts in the inner and outer disc \citep{Chiappini2015a}. Again, a thick-disc/bulge component with a more complex star-formation history than predicted by standard models might explain this observation.
\end{enumerate}

In summary, we have demonstrated that the CoRoGEE sample is well-suited for the purpose of reconstructing the chemical enrichment history of the Milky Way disc. This first study will be followed by an investigation that focusses on exploring the detailed multi-element abundance patterns provided by APOGEE. It will be based on an analysis of newly reduced CoRoT light curves, resulting in more accurate seismic parameters, and will also include data from the CoRoT long run in the LRa02 field.

From the mid-term perspective, the CoRoGEE dataset can be viewed as a pathfinder and complementary dataset to the massive surveys that the {\it Kepler}-2 mission (K2; \citealt{Howell2014}) is currently conducting.
The K2 Galactic Archaeology Program \citep{Stello2015} will deliver seismic parameters for thousands of red giants in ten fields along the ecliptic plane, and, combined with the legacy of CoRoT as well as the original {\it Kepler} mission, will enable further improvements in the coverage of the Galactic disc with solar-like oscillating red giants.

\bibliographystyle{aa}
\bibliography{FA_library}

\begin{acknowledgements}
FA would like to dedicate this work to the memory of Prof. Angelo Cassatella$^{\dagger}$. He also thanks E. C. Herenz for stimulating discussions and critical thoughts that accompanied him during the past months. TSR acknowledges support from CNPq-Brazil. BM, FB, RP and RAG acknowledge financial support from the ANR program IDEE Interaction Des \'Etoiles et des Exoplan\`etes. JM acknowledges support from the ERC Consolidator Grant funding scheme (project STARKEY, G.A. No. 615604). LG and TSR acknowledge partial support from PRIN INAF 2014 - CRA 1.05.01.94.05. TM acknowledges financial support from Belspo for contract PRODEX GAIA-DPAC. AEGP, CAP, DAGH, and OZ acknowledge support provided by the Spanish Ministry of Economy and Competitiveness (MINECO) under grants AYA2014-56359-P, RYC-2013-14182, and AYA-2014-58082-P. TCB acknowledges partial support from grant PHY 14-30152 (Physics Frontier Center / JINA-CEE) awarded from the U.S. National Science Foundation. SaM acknowledges support from the NASA grant NNX12AE17G. SzM has been supported by the J{\'a}nos Bolyai Research Scholarship of the Hungarian Academy of Sciences. The research leading to the presented results has received funding from the European Research Council under the European Community's Seventh Framework Programme (FP7/2007-2013)/ERC grant agreement No. 338251 (StellarAges).\\

The CoRoT space mission, launched on December 27 2006, was developed and 
operated by CNES, with the contribution of Austria, Belgium, Brazil, ESA (RSSD and Science Program), Germany and Spain. This research has made use of the ExoDat Database, operated at LAM-OAMP, 
Marseille, France, on behalf of the CoRoT/Exoplanet program.\\

Funding for the SDSS-III Brazilian Participation Group has been provided by the Ministério de Ciência e Tecnologia (MCT), Funda\c{c}\~{a}o Carlos Chagas Filho de Amparo à Pesquisa do Estado do Rio de Janeiro (FAPERJ), Conselho Nacional de Desenvolvimento Científico e Tecnológico (CNPq), and Financiadora de Estudos e Projetos (FINEP). Funding for SDSS-III has been provided by the Alfred P. Sloan Foundation, the Participating Institutions, the National Science Foundation, and the U.S. Department of Energy Office of Science. The SDSS-III web site is \url{http://www.sdss3.org/}. SDSS-III is managed by the Astrophysical Research Consortium for the Participating Institutions of the SDSS-III Collaboration including the University of Arizona, the Brazilian Participation Group, Brookhaven National Laboratory, Carnegie Mellon University, University of Florida, the French Participation Group, the German Participation Group, Harvard University, the Instituto de Astrofisica de Canarias, the Michigan State/Notre Dame/JINA Participation Group, Johns Hopkins University, Lawrence Berkeley National Laboratory, Max Planck Institute for Astrophysics, Max Planck Institute for Extraterrestrial Physics, New Mexico State University, New York University, Ohio State University, Pennsylvania State University, University of Portsmouth, 
Princeton University, the Spanish Participation Group, University of Tokyo, 
University of Utah, Vanderbilt University, University of Virginia, University 
of Washington, and Yale University.
\end{acknowledgements}

\Online
\begin{appendix}
 
\section{PARAM sanity checks}\label{sane}

\subsection{Seismic vs. spectroscopic gravities}
Figure \ref{aspcapseismo} shows a comparison between seismic and 
(calibrated) ASPCAP $\log g$ as a function of effective temperature. 
An immediate result is that while asteroseismology provides an 
accurate benchmark for spectroscopic gravities, spectroscopy 
may serve as an important cross-check for the determined asteroseismic 
parameters, especially for fainter stars. 
By requiring that the difference in $\log g$ not be too large, we are 
able to sort out potentially flawed seismic (or spectroscopic) parameters. 
For DR12, the ASPCAP gravities were calibrated using seismic gravities from 
{\it Kepler} \citep{Holtzman2015}. An analysis of APOKASC stars with 
known evolutionary status demonstrated that for RGB stars that have 
not yet entered the helium-burning phase, the offset between seismic and 
spectroscopic gravity is larger than for red-clump (RC) stars. Hence, one would 
ideally use two different calibration relations for the RC and RGB stars.
In the meantime, ASPCAP provides a $\log g$ calibration only for RGB stars, 
while a calibration for RC stars is reported in a separate catalogue 
\citep{Bovy2014b}.
The temperature dependence of the gravity offset also reflects the bias imposed 
by the adopted calibration relation: at lower temperatures (on the upper RGB), 
the systematic discrepancy vanishes.

\subsection{Scaling relations vs. grid-based results}

The concordance between the results obtained with PARAM and 
from the direct method has already been mentioned in 
\citet{Rodrigues2014}, who used PARAM to estimate masses, radii, and gravities 
for the APOKASC sample. In the direct method, the quantities mass, radius, and 
gravity are calculated through seismic scaling relations (which involve seismic global parameters, and $T_{\mathrm{eff}}$, but no information on metallicity or stellar models).

\begin{figure*}\centering
\includegraphics[width=.99\textwidth]
{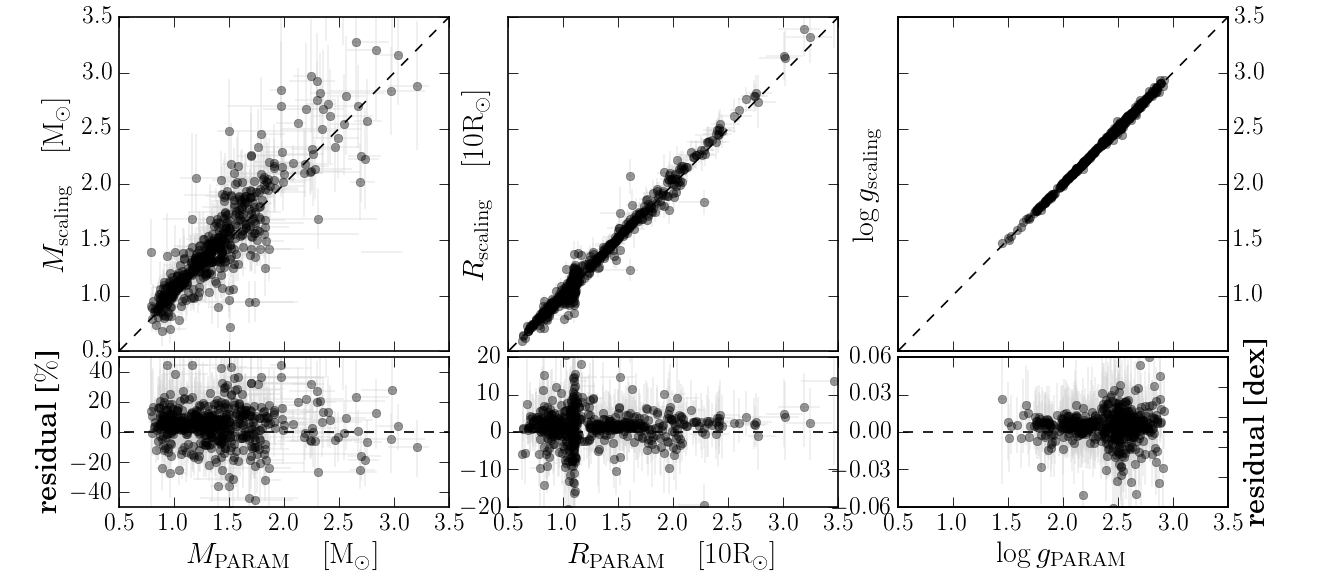}
 \caption{Comparison of our PARAM results for mass ({\it left panel}), radius 
({\it middle}), and surface 
gravity ({\it right}) with the results obtained using the direct method 
(scaling relations). 
Compare also Fig. 4 of \citet{Rodrigues2014}.}
\label{gridvsscale}
\end{figure*}

Figure \ref{gridvsscale} presents the comparison of the two methods for our sample. 
The resulting mean differences and rms scatter are $(5.3\pm13.7)\%$ in mass, 
$(1.3\pm5.1)\%$ in radius, and $0.005\pm0.012$ dex $[0.2\pm0.5)\%]$ in $\log 
g$, comparable to what was reported by \citet{Rodrigues2014} for APOKASC.

\subsection{Comparison with extinction maps}\label{ext}

\begin{figure}\centering
\includegraphics[width=.49\textwidth]
{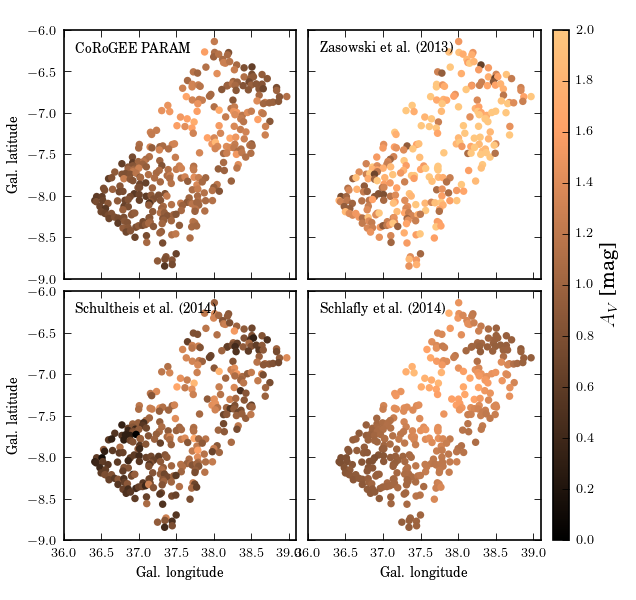}
 \caption{Comparison of our derived individual $A_V$ extinction values for 
stars in the LRc01 field with extinction estimates derived by other (mostly 
independent) methods.}
\label{extinction_Lrc01}
\end{figure}
 
\begin{figure}\centering
\includegraphics[width=.49\textwidth]
{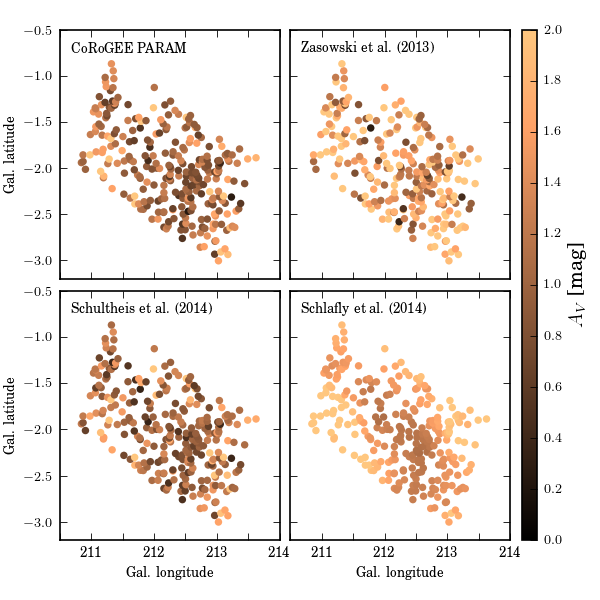}
 \caption{Same as Fig. \ref{extinction_Lrc01}, now for the LRa01 field.}
 \label{extinction_Lra01}
\end{figure}

\begin{figure*}\centering
\includegraphics[width=.99\textwidth]
{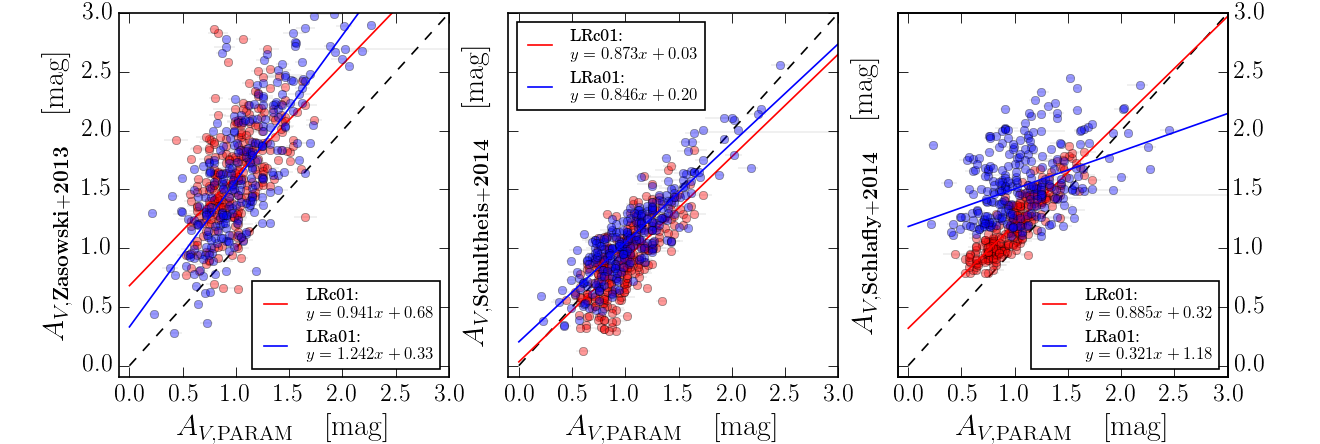}
 \caption{Comparison of our PARAM extinctions with the results obtained 
by the RJCE method (\citealt{Zasowski2013}; {\it left panel}), isochrone 
matching (\citealt{Schultheis2014}; {\it middle panel}), and the Pan-STARRS1 dust 
maps of \citet{Schlafly2014}. As before, stars in LRa01 are plotted in blue, 
while LRc01 stars are plotted in red. The corresponding robust linear fits 
(using a Huber loss function; see, e.g., \citealt{Ivezic2013}) are shown as solid lines, with the fit coefficients indicated in each panel. 
}
\label{Avcomp}
\end{figure*}

Another check is provided by Figs. \ref{extinction_Lrc01} and 
\ref{extinction_Lra01} which show $A_V$ extinction 
maps for the sample stars in the two CoRoT fields, and compare these results 
to the maps obtained using other methods: The Rayleigh-Jeans colour 
excess (RJCE) method \citep{Majewski2011, Zasowski2013}, the 
isochrone-matching method presented in \citet{Schultheis2014}, and the 
2D dust extinction maps derived from Pan-STARRS1 photometry \citep{Schlafly2014}. 
A quantitative comparison between our results and these literature methods, 
together with empirical fitting formulae for each extinction scale, is 
presented in Fig. \ref{Avcomp}. In summary, we can say the following:

\begin{itemize}
 \item The RJCE method \citep{Majewski2011} relies on the fact that the intrinsic NIR -- mid-IR colours (e.g., $H_{\mathrm{2MASS}}-W2_{\mathrm{WISE}}$) of a star depend very little on the spectral type, and therefore the observed minus intrinsic colour provides a measurement of the amount of dust in the sightline of an observer. 
 The comparison with the extinction values calculated using this recipe (which was used for APOGEE targeting; \citealt{Zasowski2013}) shows that -- assuming a particular extinction law \citep{Nishiyama2009} -- RJCE overpredicts the amount of $V$-band extinction in both LRa01 and LRc01 by about 0.5 mag. Of course, as APOGEE operates in the $H$ band ($A_H/A_V\approx1/6$), this systematic difference is of minor importance for APOGEE targeting purposes. However, our comparison shows that, when computing distances to APOGEE field stars (e.g., \citealt{Anders2014, Santiago2016}), we should be cautious in using the targeting extinction values; in particular, distant low-latitude stars will be assigned systematically greater distances.
 \item The isochrone-based method of \citet{Schultheis2014}, tailored to quantifying 3D extinction towards the Galactic bulge, yields slightly lower extinction values than our method; there is only a minor zero-point offset of about 0.05 mag in the extinction scale with respect to PARAM (middle panel of Fig. \ref{Avcomp}). When this effect is calibrated out, the rms scatter around the mean relation is about 0.2 mag in both fields.
 \item \citet{Schlafly2014} used multi-band photometry star-counts from Pan-STARRS1 \citep{Kaiser2010} to create a 2D $E(B-V)$ reddening map, quantifying integrated interstellar extinction at heliocentric distances of 4.5 kpc.
 The resolution at low Galactic latitudes is typically $7\arcmin$ and the systematic uncertainty in $E(B-V)$ around 0.03 mag.
 Our results show that while the overall amount of extinction for the bulk of the CoRoGEE sample is reproduced by the Pan-STARRS maps, the relation between our extinction estimates and those derived from Pan-STARRS is dominated by considerable scatter, especially in the LRc01 field. 
 This result is expected, as most of our stars lie within the 4.5 kpc boundary, some even closer than 1 kpc from the Sun.
 \item Not shown in Fig. \ref{Avcomp} is the comparison of our results with the classical 2D extinction SFD maps of \citet*{Schlegel1998}, as for Galactic astronomy purposes, they are surpassed in accuracy by the maps of \citet{Schlafly2014}. It is worth mentioning, however, that in the LRc01 field (only $7^{\circ}$ off the Galactic 
 plane) our method agrees well with the SFD maps, also on a star-by-star level; we find a very tight relation between $A_{V, \mathrm{SFD}}$ and $A_{V, \mathrm{PARAM}}$ in this field, with an rms scatter of $\sim0.15$ mag. This suggests that the extinction in this field is likely to be dominated by a nearby foreground cloud (as also visible in the WISE image of Fig. \ref{fieldmaps}).

 In the LRa01 field, however, the situation is not as favourable: The SFD maps overpredict the extinction in LRa01 by more than one magnitude on average, and the correlation with the PARAM results is marginal. This finding agrees with previous studies close to Galactic plane (e.g., \citealt{Peek2010, Schlafly2011}), and might be explained by significant additional amounts of dust beyond the bulk of the CoRoGEE stars (e.g., the Galactic warp).
\end{itemize}

\section{Released data}\label{catapp}
In Table B.1, we shortly summarise the contents of this first set of CoRoGEE data that is released through the CDS Vizier Catalogue Service\footnote{\url{vizier.u-strasbg.fr/viz-bin/VizieR}}. 

The present CoRoT-APOGEE dataset contains a large amount of information (206 columns) on the 606 successfully observed stars. In addition to the measurements derived directly from APOGEE and CoRoT observations, we include photometry from OBSCAT, APASS, SDSS, 2MASS, and WISE, information from the EXODAT archive, stellar parameters, distances and extinctions from PARAM and/or seismic scaling relations, cross-matches to the APOGEE DR12 RC catalogue \citep{Bovy2014b}, the UCAC-4 catalogue \citep{Zacharias2013}, and additional information on the kinematics of the stars.

\begin{longtab}
\centering
{\footnotesize
\include{corogee_tablecols_long2}
}
\label{cataloguetab}
\end{longtab}

\end{appendix}

\end{document}